\documentclass[acmlarge]{acmart}
\usepackage{mdframed}
\usepackage{booktabs} 
\usepackage{subfiles}
\usepackage{booktabs}
\usepackage{multirow}
\usepackage{wrapfig}
\usepackage{color}
\usepackage{soul}
\usepackage{comment}
\usepackage{hyperref}
\usepackage{subcaption}
\usepackage[nohyperlinks, printonlyused, withpage, smaller]{acronym}
\usepackage[inline]{enumitem}
\definecolor{mycolor}{RGB}{0,112,192}
\renewcommand\hl[1]{#1} %
\AtBeginDocument{%
  \providecommand\BibTeX{{%
    \normalfont B\kern-0.5em{\scshape i\kern-0.25em b}\kern-0.8em\TeX}}}

\setcopyright{acmcopyright}
\copyrightyear{2021}
\acmYear{2021}
\acmDOI{00.000/0.0}

\acmJournal{IMWUT}
\acmVolume{4}
\acmNumber{3}
\acmArticle{79}
\acmMonth{11}
\acmPrice{15.00}
\acmDOI{10.1145/3411813}

\begin{document}

\title{Investigating the Effects of Mood \& Usage Behaviour on Notification Response Time}


\author{Judith Simone Heinisch$^\ast$}
\affiliation{%
  \institution{University of Kassel}
  \department{Department for Communication Technology}
  \streetaddress{Wilhelmsh{\"o}her Allee 73}
  \city{Kassel}
  \state{Hessen}
  \postcode{34121}
  \country{Germany}}
  
\author{Nan Gao$^\ast$}
\affiliation{%
  \institution{RMIT University}
  \department{School of Computing Technologies}
  \streetaddress{}
  \city{Melbourne}
  \state{VIC}
  \postcode{3000}
  \country{Australia}}

\author{Christoph Anderson}
\affiliation{%
  \institution{University of Kassel}
  \department{Department for Communication Technology}
  \streetaddress{Wilhelmsh{\"o}her Allee 73}
  \city{Kassel}
  \state{Hessen}
  \postcode{34121}
  \country{Germany}}
  
\author{Shohreh Deldari}
\affiliation{%
  \institution{RMIT University}
  \department{School of Computing Technologies}
  \streetaddress{}
  \city{Melbourne}
  \state{VIC}
  \postcode{3000}
  \country{Australia}}
  
\author{Klaus David}
\affiliation{%
  \institution{University of Kassel}
  \department{Department for Communication Technology}
  \streetaddress{Wilhelmsh{\"o}her Allee 73}
  \city{Kassel}
  \state{Hessen}
  \postcode{34121}
  \country{Germany}}

\author{Flora Salim}
\affiliation{%
  \institution{University of New South Wales (UNSW)}
  \department{School of Computer Science and Engineering}
  \streetaddress{}
  \city{Sydney}
  \state{UNSW}
  \postcode{1466}
  \country{Australia}}
  
\thanks{$^\ast$These authors contributed equally to this work.}

\newacro{esm}[ESM]{Experience Sampling Method}
\newacro{sam}[SAM]{Self-Assessment Manikin}

\newacro{iv}[IV]{Independent Variable}
\newacro{dv}[DV]{Dependent Variable}
\newacro{lsdv}[LSDV]{Least Squares Dummy Variable Model}
\newacro{svm}[SVM]{Support Vector Machine}
\newacro{knn}[KNN]{K-Nearest Neighbors}
\newacro{rf}[RF]{Random-Forests}
\newacro{nb}[NB]{Multinomial Naive Bayes}
\newacro{rrr}[RRR]{Relative Risk Ratio}

\newacro{bvp}[BVP]{Blood Volume Pulse}
\newacro{eda}[EDA]{Electrodermal Activity}
\newacro{st}[ST]{Skin Temperature}
\newacro{hr}[HR]{Heart Rate}
\newacro{hrv}[HRV]{Heart Rate Variability}
\newacro{ibi}[IBI]{Inter Beat Interval}
\newacro{ppg}[PPG]{Photoplethysmography}
\newacro{acc}[ACC]{Accelerometer}
\newacro{scl}[SCL]{Skin Conductance Level}
\newacro{scr}[SCR]{Skin Conductance Response}

\newacro{cpd}[CPD]{Change Point Detection}

\newacro{tf-idf}[TF-IDF]{Term Frequency-Inverse Document Frequency}
\newacro{tf}[TF]{Term Frequency}
\newacro{cv}[CV]{Count Vectorization}
\newacro{w2v}[W2V]{Word2Vec}
\newacro{iqr}[IQR]{interquartile range}

\newacroplural{esm}[ESMs]{Experience Sampling Methods}
\newacroplural{iv}[IVs]{Independent Variables}
\newacroplural{dv}[DVs]{Dependent Variables}
\newacroplural{lsdv}[LSDVs]{Least Squares Dummy Variable Models}
\newacroplural{svms}[SVMs]{Support VectorMachines}

\renewcommand{\shortauthors}{Anonymous et al.}

\begin{abstract}
Notifications are one of the most prevailing mechanisms on smartphones and personal computers to convey timely and important information. Despite these benefits, \hl{smartphone} notifications demand individuals' attention and can cause stress and frustration when delivered at \hl{inopportune} timings. This paper investigates the effect of individuals' smartphone usage behavior and mood on notification response time. We conduct an \textit{in-the-wild} study with more than \hl{$18$} participants for five weeks. Extensive experiment results show that the proposed regression model is able to accurately predict the response time of \hl{smartphone} notifications using current user's mood and physiological signals. We explored the effect of different features for each participant to choose the most important user-oriented features in order to to achieve a meaningful and personalised notification response prediction. On average, our regression model achieved over all participants an MAE of $0.7764$~\hl{ms} and RMSE of $1.0527$~\hl{ms}. We also investigate how physiological signals (collected from E4 wristbands) are used as an indicator for mood and discuss the individual differences in application usage and categories of smartphone applications on the response time of notifications. Our research sheds light on the future intelligent notification management system. 
\end{abstract}



\maketitle

\section{Introduction}


Smartphones, laptops, and desktop computers play an important role in human everyday lives. These devices frequently send people notifications such as emails, messages, news, application update information, etc. Inappropriate interruptions can lead to user annoyance and anxiety \cite{bailey2001effects}, decrease productivity \cite{renaud2006you} and task performance \cite{bailey2000measuring}, or affect emotional state \cite{bailey2001effects}. For instance, Perlow et al. \cite{perlow1999time} found that the software engineers in a high technology company had difficulties meeting deadlines due to frequent interruptions.
These examples highlight the importance of interruption management as an emerging field of research to reduce distractions.

Human attention is a finite resource. When people perform a task, an interruption can split the attention resource into two interactive tasks \cite{kim2015sensors}. People need to estimate whether the benefits of the interrupted interaction are high enough to offset the loss of attention in the original task. Different actions can be taken to deal with interruptions, such as ignoring, postponing the processing to a more convenient time, or immediately resolving the interruptions. Different measures may delay resuming the original task and reduce the task performance to varying degrees \cite{trafton2003preparing}.

\textit{Receptivity} refers to a user's reaction to an interruption which may indicate both the level of interruptibility of the user and their experience of the interruption \cite{Fischer2010}. In some cases, even though the notification is interruptive, the user can still be receptive to the notification. Previous studies have shown that users' receptivity to notifications is influenced by many factors: (1) informational qualities of the notifications, e.g. interest, entertainment, relevance and actionability \cite{Fischer2010}; (2) mobile usage, such as the time of the interruption and the type of app pushing the notification \cite{Shirazi2014,Fischer2010} ; (3) demographics, such as personality traits \cite{Yuan2017}; (4) personal dynamics, such as location \cite{exler2016preliminary}, transitions between activities \cite{Ho2005} and social roles \cite{Anderson2016}. 

\hl{However, we propose a system in a real-world scenario to help manage the automatic pop-up notifications of frequently used smartphone applications, which has not been attempted by other researchers before.} Users' receptivity varies based on physical, psychological, and affective conditions , and the accuracy of existing systems in addressing these conditions is still relatively low \cite{Mehrotra2016a}. The difficulty of including these conditions can be explained by an example : Users may get annoyed (psychological) if an email from their boss suddenly pops up while they are concentrating on writing a paper and are in a state of 'flow' (physical). However, it is not clear how the user would feel (affect) if this email notifications were postponed. On the one hand, they may be relieved at not being disturbed, but on the other hand, it could cause stress if they were waiting for important information to help them with a problem they are experiencing.

Therefore, in this research, we aim to bridge this gap and conduct an \textit{in-the-wild} study in a \textit{multi-device} setting to collect user behaviour along with contextual information, interruptibility, receptivity, mood and social roles from more than $18$ participants during five weeks. We have designed two applications, \textit{Balance for desktop} and \textit{Balance for Android}, which use \textit{ESM} \cite{Hektner2007} to capture users' interruptibility preferences, user behavior, and mood toward the notifications \hl{on smartphones}. Meanwhile, participants are asked to wear a wristband to record their physiological signals (\ac{eda}, \ac{bvp}, and \ac{st}). 
We summarize the main contributions as follows:

\begin{itemize}
    \item We conduct an in-situ study with $27$ participants \hl{over a five-week} period. In total, we \hl{collected 42,270 notifications with 3,236} ESM responses and more than \hl{5,920} hours of physiological signals from Empatica E4 wristbands. To the best of our knowledge, this is the most heterogeneous and diverse data set collected \textit{in-the-wild} to study the notification response behaviour of users.
    
    \item  We explore diverse notification response behaviours of different participants and investigate the relationships between multiple factors (\hl{e.g.} mood \hl{and} apps) and notification response \hl{times}. We found a statistically significant correlation between response time and in-use \hl{apps}.
    
    
    \item We \hl{conduct} extensive experiments to predict the notification response time for each participant. The experiment results show that the proposed model (\textit{Bayeian Rdige} Regressor) achieves high prediction performance (\hl{MAE = 0.7764 and RMSE = 1.0527}). We \hl{then} derive the most useful features for each participant to achieve a meaningful and personalised \hl{prediction of} notification. 
    \item We investigate how the mood-related features improve the prediction performance by \hl{utilising} the ESM responses and physiological signals (\hl{e.g. EDA and HR}). We further discuss various factors affecting the prediction performance, such as the individual differences and categories of apps.
\end{itemize}

The investigation of mood, physiological signals, and usage behaviour on users' receptivity to notifications \hl{on the smartphone} leads to new sights to the future notification management system. The remainder of the paper is as follows. Section \ref{related_work} introduces related works of interruptibility management, receptivity and popular mood sensing approaches. Section \ref{methodology_study} describes the data collection procedures, including participant recruitment, applications designed for data collection, \ac{esm} questionnaires, and collected data types. Then we introduce the pre-processing techniques and extracted features in Section \ref{sec:method}. In Section \ref{sec:EDA}, we analyze notification response behaviours across different participants in various scenarios. Section \ref{sec:experiment} shows the experimental results for predicting the notification response time, and Section \ref{sec:implications} lists the limitations and implications of this research. Finally, we summarize our findings in Section \ref{sec:conclusion} and indicate the potential directions in future research.




\section{Background}
\label{related_work}

\subsection{Interruptibility Management}
We considered the current state of the art regarding response times. In this paper, we define response time as the time that elapses between \hl{receiving a notification and} opening \hl{the} corresponding \hl{app}.

Okoshi et al. \cite{Okoshi2019} \hl{presented} a system \hl{to detect} opportune moments for interruptions based on click rate gain \hl{using} mobile sensing and \hl{ML} methods. They \hl{calculated} the user's click response \hl{times} by measuring the time between a notification's arrival and the response to \hl{the notification, i.e. click on the notification.} This data \hl{was logged} along with \hl{contextual} information from the smartphone and \hl{the data were} evaluated. A trained linear regression model \hl{then identified} whether \hl{a} moment \hl{in time was} an opportune moment \hl{to display a notification based} on the extracted features. The adaptive notification component \hl{then delayed} the presentation of notifications to the user until \hl{an opportune} moment \hl{was} detected. \hl{This breakpoint-based notification scheduling system} resulted in increased click rates and quicker responses from \hl{users}.

Saikia et al. \cite{Saikia2017}, developed an optimization process to reduce the reaction time and increase the opening rate of notifications for a mobile news application. Like Okoshi et al. \cite{Okoshi2019} they defined the response time as the time between receiving and opening notification and gathered additional context data (such as the category of notification, time of the day, location, etc.). Also, the notification opening rate, which is similar to the click rate \cite{Okoshi2019}, was used, to optimize the opening rate and minimize the response time. With their framework, Saikia et al. reached a decreased reaction time by 13,3\% and improved the opening rates by 65.24\%.

 Westermann et al. \cite{Westermann2016} have studied the significance of the context factor time, regarding the time of the day and weekdays, on receptivity to notifications based on android smartphones. For this, they sent advertisement notifications about popular brochures. The response time was set as the time between receiving a notification and opening the app. Results exhibit notable variations in response times and notification-triggered app launch numbers for weekdays and different time slots.

The authors \cite{Visuri2019} developed an application to log self-reported data on the significance of notification contents, notification source, and delivered context to analyze the relationship between notification interaction choices and content importance. They also collected data such as contextual information, notification content, time of delivery, and whether the user clicked on the notification or dismissed it. Based on user ratings of past notifications, contextual data, and semantic analysis, a predictive Machine Learning model is created to predict whether future notifications are useful or not. The results showed that considering only interactions like click or dismiss ratios does not suffice to classify the importance of notification as users mostly tend to ignore notifications irrespective of their importance. Using semantic analysis of notification content enhances the accuracy of the prediction model.


The paper by Fortin et al.\cite{fortin2019} highlights the correlation between skin conductance response (SCR) and the prediction of the perception of smartphone notifications. To study the impact of user activity on the determined signals, the participants were asked to perform an inactive (watch a wildlife documentary) and active (solve paper mazes) task during the measurement. They were then directed to note the stimulus (sound or vibration of the phone) that made them perceive the notification and press the corresponding buttons on a Pebble smartwatch placed next to them. The experiments showed that notifications perceived because of their tactile properties (vibration) stimulated larger skin conductance responses and SCR with higher amplitudes compared to those perceived through auditory properties (sound). A logistic regression model was trained to examine if a perception prediction method using skin conductance could aid notifications, including the smartphone's ringer mode as a predictor variable. This model successfully identified perception in 75\% of true cases when participants perceived the notification and 38\% of missed notifications.

\subsection{Receptivity}
In \cite{Mehrotra2016a}, Mehrotra et al. study the effects of cognitive and physical contextual information on individuals' receptivity towards notifications. Lee et al. \cite{Lee2019} investigate the correlation of individuals' relationships to contacts and contextual descriptors on the receptivity to mobile instant-messaging notifications. The authors find that contextual descriptors such as the engagement in activities are more descriptive than the relationship to contacts when predicting receptivity.

In this study \cite{Mehrotra2016}, the authors have investigated the factors that make a smartphone notification disruptive and its impact on the response time. An Android app called "my phone and me" was created. The application uses Android's Notification Listener Service to access notifications and Google's Activity Recognition API and ESSensorManager to receive context info. Apart from this, the app also triggers questionnaires every 4 hours between 8 am and 8 pm.  Reaction time is considered as the time from the notification arrival till the time it was reacted upon. The modes of identifying notification (ringer or vibration) and the user's personality traits were also noted. The results showed that high-priority notifications were responded much faster, whereas those from less frequent contacts were responded too late. Also, notifications are considered highly disruptive, when the user was performing a task and least disruptive before starting a new task or being idle.

Züger et al. \cite{Zueger2018} predicted the interruptibility of 13 software developers on computer interaction, heart-, sleep-, and physical activity-related data. They found that the interaction with a computer gives more information about interruptibility than biometric data. However, using both data outperformed the results ahead.

\subsection{Mood Sensing Approaches}
Before we turn to papers in the area of Attention Management, let us give a definition of the term mood, which is frequently used in this paper.
Mood is a diffuse affective state that describes an individual's subjective feeling over a long time. Unlike emotions, mood lasts over hours or days, and the intensity is usually low. Most of the time, we cannot assign a specific trigger to our mood, or to name a reason. Nonetheless, mood influences our behavior and experiences \cite{scherer2005emotions}.

Changes in activities, moods and behavior of users provide valuable insights on providing context-aware services and minimizing unwanted interruptions. According to recent researches in psychology frequency of changes or the rate of instability in different characteristics can affect the interruptability of users \cite{couffe2017failures}. In the field of Attention Management, different consequences were investigated. Among others, the influence of interruptions on our mood.
Zijlstra \cite{Zijlstra1999}, for example, identified interruptions resulting in negative emotions. The counterpart is our mood as an internal stimulus, which results from our insights, and influences our interruptibility \cite{Miyata1986,Dabbish2011}. Therefore, emotions and stress are not only consequences of interruptions but also influencing our interruptibility.

Yuan et al. \cite{Yuan2017} proposed not only using personality traits to group similar users, but they also considered different contextual information such as location, changes in the state of the user, time, transition state, current activity, and mood to predict reactions to interrupts and also interruptibility intensity.
Kaur et al. \cite{Kaur2020} developed a real-time system recommending during opportune moments transitions and breaks while not disrupting people during their focused states. They evaluated their system with a combination of emotions (classified user's facial expression), productivity (daily task list), and self-reports. Using personalized models, they achieved an accuracy of 86\% and 77\% for predicting opportune moments for transitions and breaks, respectively.

Khan et al. \cite{Kahn2020} propose a new approach for Automated Mood Recognition (AMR) in a smart office environment, which reduces computational requirements by requiring fewer mood models. This is done by clustering physiological signals by groups of people who sense emotions in the same way. They used machine learning models for classification and regression, which are trained based on the extracted features of users in common perception clusters recognizing the mood. Eight different categories of moods are recognized, each with three different levels denoting low, medium, and high intensities. The proposed approach seems to be a trade-off between the requirement of a large number of personalized mood models and the insufficient performance of generalized models for AMR.  Results show respective average F1 scores of 0.76 and 0.79 for perception clusters and personalized-based AMR.

\subsection{Relevance to our approach}

Current approaches in the field of attention management already concentrate on notifications and their consequences on human behavior and well-being. It is already known that receiving notifications can negatively impact on our mood and trigger stress. Likewise, the reverse case had shown that our mood influences our behavior towards notifications and our interruptibility. We want to go one step further and look at the effects of our mood on the response time to notifications. For this purpose, we extend the current state of research by adding physiological signals to the moods captured via ESM. We want to identify whether the mood directly affects the response time. Using individual regression models, we predict the receptivity of each user.
\section{Data Collection}
\label{methodology_study}
In this section, we describe the design and data collection of our \textit{in-the-wild} study. \hl{First, we give a general overview. Further, we provide some insights about the participants we measured, before we explain the applications and collected data in detail.}

\subsection{Overview}
\label{balance}

\begin{figure}[t]
    \centering
    \includegraphics[scale=.6]{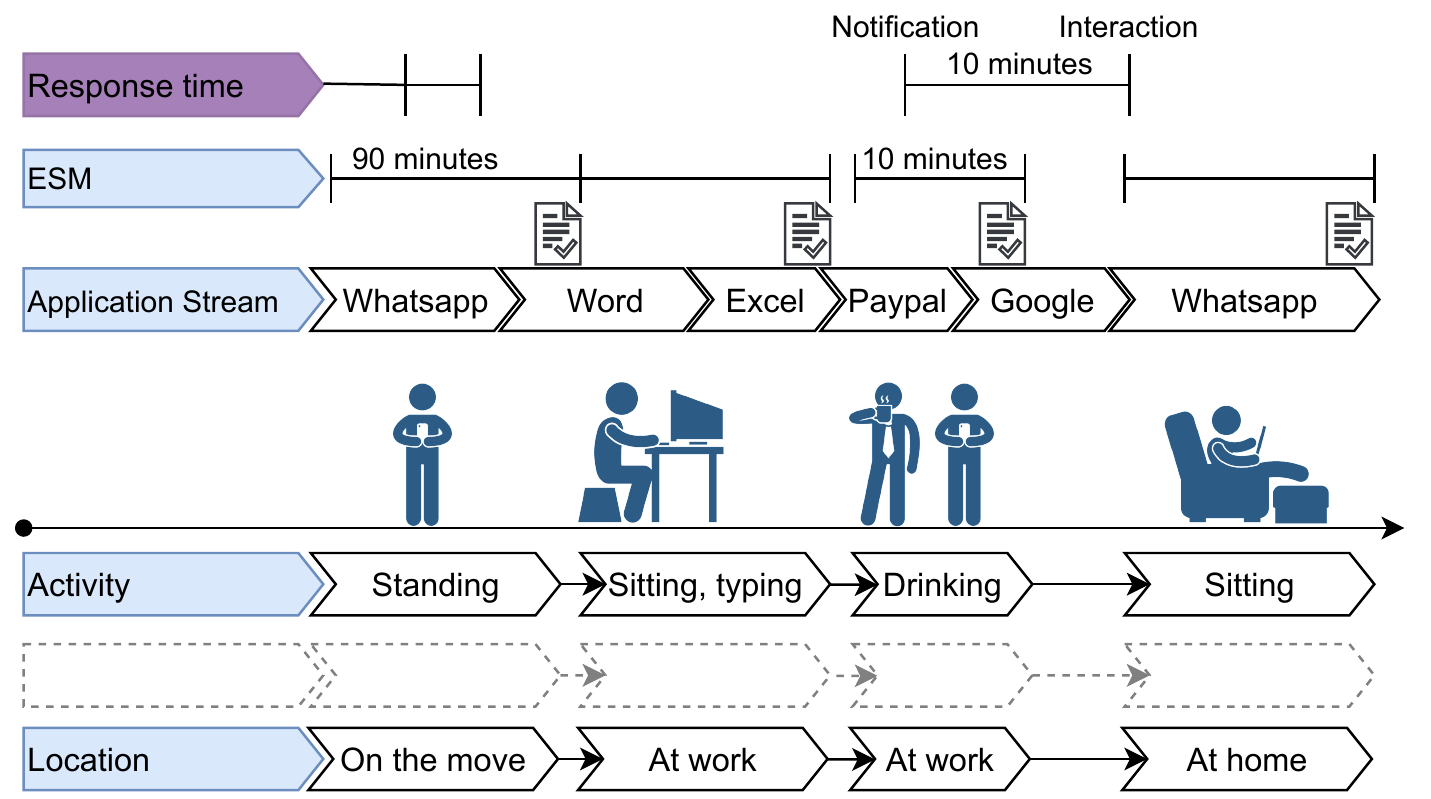}
    \caption{Study design to capture interruptibility, mood and response time.}
    \label{fig:study_design}
\end{figure}

We performed an \textit{in-the-wild} study to gather user behavior regarding \hl{smartphone} notification arrival, response time, combined with contextual information and mood via \hl{Android smartphones, desktop computers, and physiological signals}. By advertising our study on our websites and networks, we acquired $27$ participants for our field study. The data collection began at the end of January 2020 and continued for five weeks. The participants were asked, to install the applications \textit{Balance for Android} (see Figure \ref{fig:android_app})  and \textit{Balance for desktop} (see Figure \ref{fig:desktop_app}) on their \hl{smartphones} and desktop computers, respectively. Both applications facilitate continuous background sensing, as well as Experience Sampling Methods (ESMs) \cite{Hektner2007}. The participants were free to additionally choose to have their physiological signals measured via an E4 wristband. This part of the measurement was coordinated and supervised by a contact person from the respective country of origin. By installing the applications or putting the E4 wristband on, the participants received information about the study and the data collected. After that, privacy protection measures and the rights of the participants were introduced (e.g., erasing their collected data on request). Before the participants were guided by a short tutorial on using the corresponding applications and handling E4, they had to give their informed consent. Our privacy department and ethics committee approved the consent forms and data collection procedures.
Our study design (see Figure \ref{fig:study_design}) called for most contextual information to be recorded in the background without the participant's input, such as running applications, physical activities, or locations.
Another part of our mixed-method approach was to present scheduled questionnaires every $90$ minutes. We asked the participants about their mood, social role, interruptibility, and the kind of task they were working on in the last $15$ minutes. Additionally, we implemented an event-based approach to show the questionnaire, which was activated after the participant interacted more than 10 minutes with their phones. These scheduled questionnaires were limited to the time between 7am and 10pm and the event-based ones had a minimum time of $30$ minutes between each other. Preventative we implemented the same limitation not to push a questionnaire after the participant received a periodic one. With these restrictions, we addressed the strain of responding to questionnaires and ensuring the quality of data \cite{vanBerkel2017,vanBerkel2019}. All used approaches are well-known in ESM based studies to capture contextual information \textit{in-situ} \cite{vanBerkel2017}.
Figure \ref{fig:study_design} depicts the design of the study, explained above.

\subsection{Participants}
\label{sec:participants_intro}
\hl{In our experiments, we focus on response time regarding smartphone notifications. Because of that, we} took $18$ of all measured $27$ participants into account \hl{($15$ male, $2$ female, $1$ diverse). The remaining $9$ had to be removed because they had not used a smartphone, we had not sufficient answers from them on the ESM questionnaires, or some technical problems affected the data collected by them.} Our participants were between $25$ and $41$ years old \hl{(mean = $31.89$ and std=$3.85$ years)} and could be acquired from five different countries on \hl{three continents, e.g. Asia, Oceania, and Europe. All participants came from a university domain comprising junior and senior scientists and technical staff members.} We found $18$ Android, $11$ Windows, and $7$ macOS users in our measurement. $15$ participants installed both, the smartphone and the desktop application,\hl{ and $12$ of them wore an E4 wristband additionally} \hl{(see Table} \ref{tab:participants}). The data was regularly transmitted to a server hosted at our university and stored in an internal database. The upload, as well as the data, were encrypted. Compared to other ESM-based studies within the field of interruptibility, the overall answering rate ($28.37 \%$) is comparable to similar studies~\cite{Pejovic2014}. So to say, $3504$ out of $12352$ questionnaires were answered.

\begin{table}[]
\label{tab:participants}
\caption{Number of users per device and gender.}
\begin{tabular}{lllll} \toprule
& \multicolumn{3}{l}{gender}    &  \\
device                                & female & male & not specified &  \\ \cline{1-4}
Smartphone only                       & 0      & 2    & 1             &  \\
Smartphone and desktop                & 0      & 3   & 0             &  \\
Smartphone, desktop, and E4 wristband & 2      & 10   & 0             &  \\ \cline{1-4}
\end{tabular}
\end{table}

\subsection{Collected data}
\label{methodology_tech}
In this section, we introduce the two applications \textit{Balance for Android} and \textit{Balance for desktop} (see Figure 2), which we implemented to capture user behavior on desktop computers and Android smartphones. The technical details and used concepts for gathering background data and experience sampling will be given. First, we focus on the implementation of the multi-platform application for \textit{Windows} and \textit{macOS}. Afterward, we will explain the \textit{Balance for Android}.


\begin{figure}[tbh]
	\begin{subfigure}[b]{0.49\textwidth}
	    \centering
		\includegraphics[scale=.06]{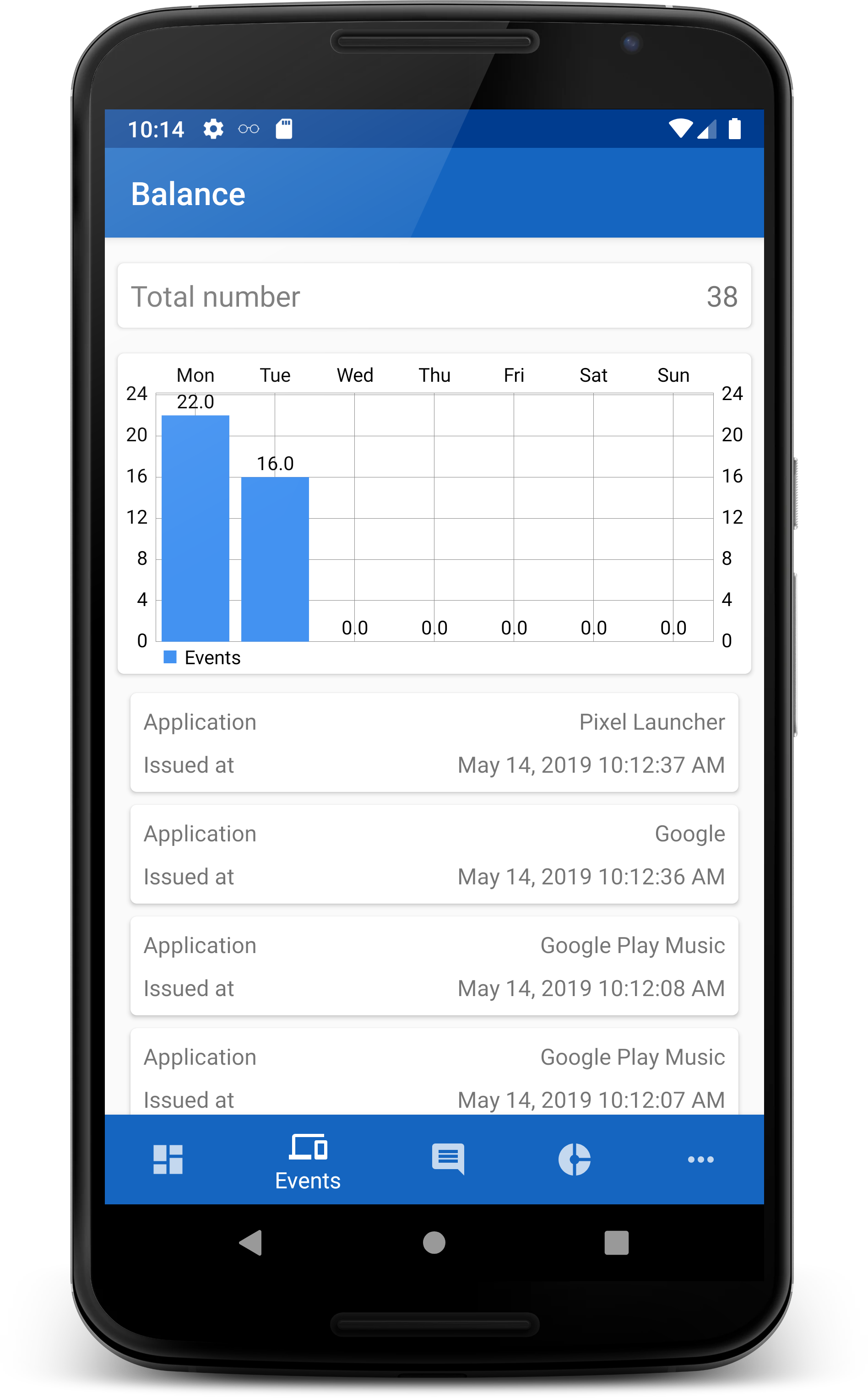}
		\caption{Balance for Android: Showing the dashboard that displays recent events that have been recorded.}
		\label{fig:android_app}
	\end{subfigure}
	\hfill
	\begin{subfigure}[b]{0.49\textwidth}
	    \centering  
		\includegraphics[scale=.16]{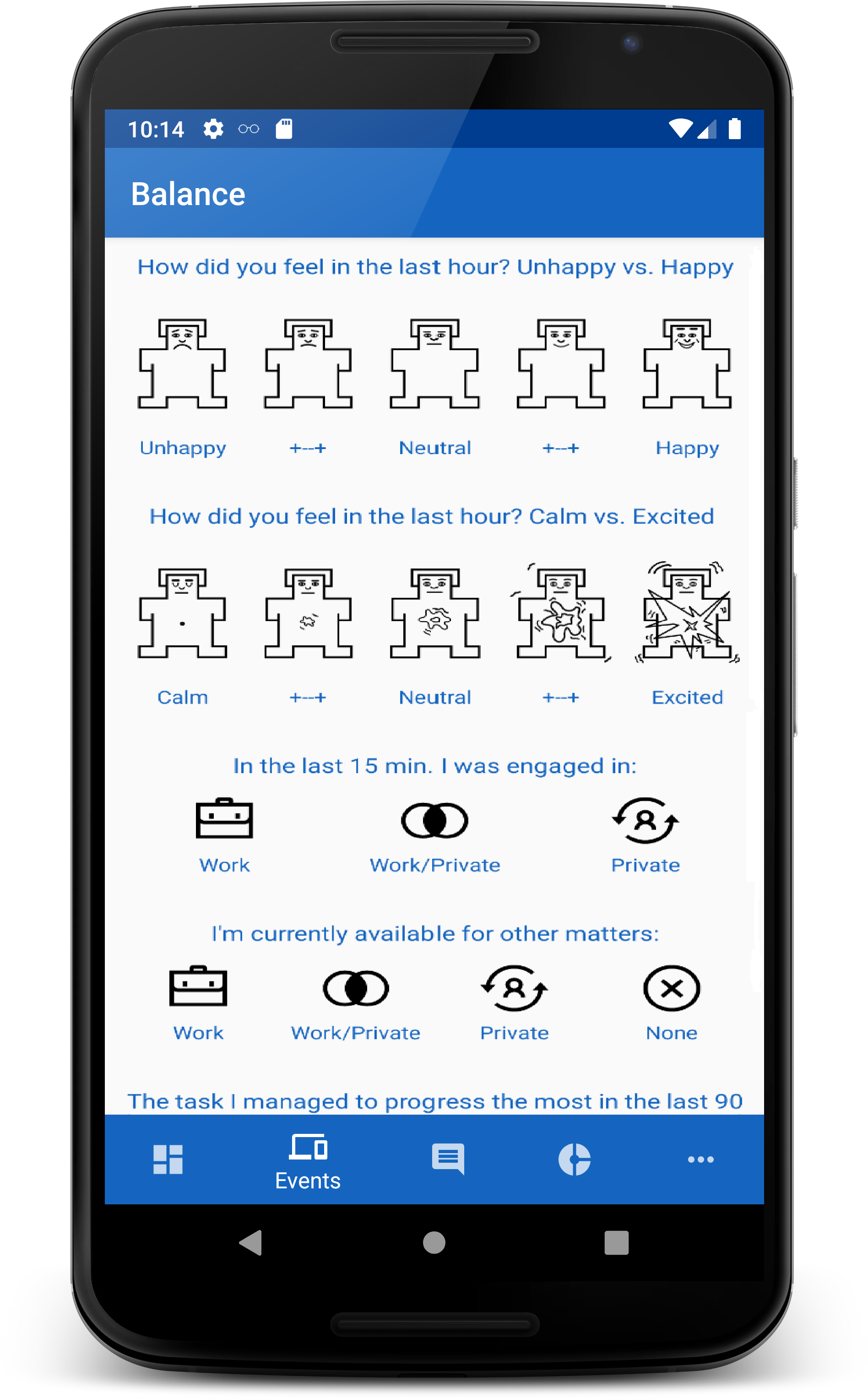}
		\caption{\hl{Screenshot of the ESM gathering mood information using SAM from Bradley and Lang.}}
		\label{fig:mood_esm}
	\end{subfigure}
	\caption{\hl{Screenshots of the smartphone application} \textit{Balance for Android} (Figure \ref{fig:android_app}).\hl{ It provided background sensing, experience sampling, and was built to capture user behavior in a multi-device setting. The application supported English and German. The mood ESM }(Figure \ref{fig:mood_esm}) \hl{used the Self-Assessment Manikin (SAM) from Bradley and Lang}  \cite{bradley1994measuring} \hl{to gather the arousal and valence state.}}
\end{figure}

\subsubsection{Balance for Windows \& macOS}

We decided to use a multi-platform application to cover the broadest possible range of users, either using Windows or macOS as an operating system. The access to foreground applications, the information given in their title bars, as well as keyboard and mouse events were provided by the libraries \textit{pywin32\footnote{See: \url{https://pypi.org/project/pywin32/}}} and the \textit{pyobjc\footnote{See: \url{https://pypi.org/project/pyobjc/}}} on Windows and macOS, respectively. Both libraries are wrappers to low-level native operating system interfaces that allow direct access to system information, peripheral devices, and functions.
Another two libraries our applications relied on are the \textit{psutil\footnote{See: \url{https://pypi.org/project/psutil/}}} and \textit{subprocess32\footnote{See: \url{https://pypi.org/project/subprocess32/}}} libraries. We used the cross-platform library Psutil to abstract information about system load and access to running processes. This includes, among other things, retrieving battery information such as the remaining charge and power states. With the \textit{subprocess32} library and native system calls, we parsed and scanned nearby Wi-Fi networks.

\subsubsection{Balance for Android}
The functionalities of the Balance for Android are very similar to the Balance for desktop computers. Analog to the desktop application, the Balance for Android also regularly uploads the recorded data encrypted to the university server. The main focus of the design was the low battery consumption, the limited resources, and the seamlessly recording of the data in the background. With the background services, we kept track of interactions with applications and notifications, location updates, and the phone's state (e.g., screen status, ringer-modes). The phone's last known location was processed by a Fused Location Provider \footnote{See: \url{https://developers.google.com/location-context/fused-location-provider/}}, an API to estimate location information. It manages the Wi-Fi, mobile communication services, and GPS while improving battery performance and resource consumption. Besides, we gathered information on physical activities by using the Google Recognition API\footnote{See: \url{https://developers.google.com/location-context/activity-recognition/}}. This API offers to report recognized physical activities and besides optimizing the battery performance. The optimized battery performance is achieved by reducing updates when the device is idle and using low-power sensors until the activity is reported.


\textbf{Applications \& Notifications}.
\label{applications_notifications}
Accessibility Services\footnote{See: \url{https://developer.android.com/reference/android/accessibilityservice/AccessibilityService}} or  Notification Listeners\footnote{See: \url{https://developer.android.com/reference/android/service/notification/NotificationListenerService}} are common methods to gather data on applications and notifications using Android, in the field of interruption management~\cite{Visuri2019,Weber2019}.
We used the Accessibility Service to gather the name and the package identifier of the used application, running in the foreground, from the smartphone. This record always happens when the window or its state changes. Another Service integrated was the Notification Listener. It intercepts the reception and removal of notifications and accesses their underlying representation. This helps us to get information like the time of arrival of the notification, the contact and group names the notification came from, or the length of the notification's content. To extract the contacts and group names, we set some applications on a white-list to process their notifications on the smartphone directly. As we were only interested in contacts, we set only popular messaging applications, like Whatsapp, Outlook, Twitter, Facebook, Microsoft Teams, Slack, or Telegram, on the list.

In order to infer the responsibility of the user and to distinguish between the notifications we asked the user about the relationship to the sender. The user could choose between the relationships \textit{family}, \textit{friend}, \textit{work}, and \textit{none}, whereby multiple naming was possible. As the senders were transmitted pseudonymized for data protection reasons, it could not be detected if a sender was named differently in different messengers or was part of a group chat. Therefore, we could not avoid sending multiple relationship questionnaires over one sender with different names. So that this additional questionnaire does not negatively influence the response rate, a minimum of correspondences with this sender was assumed before this questionnaire was triggered.

\subsubsection{Physiological data}

\begin{figure}
    \centering
    \includegraphics[width=0.33\textwidth]{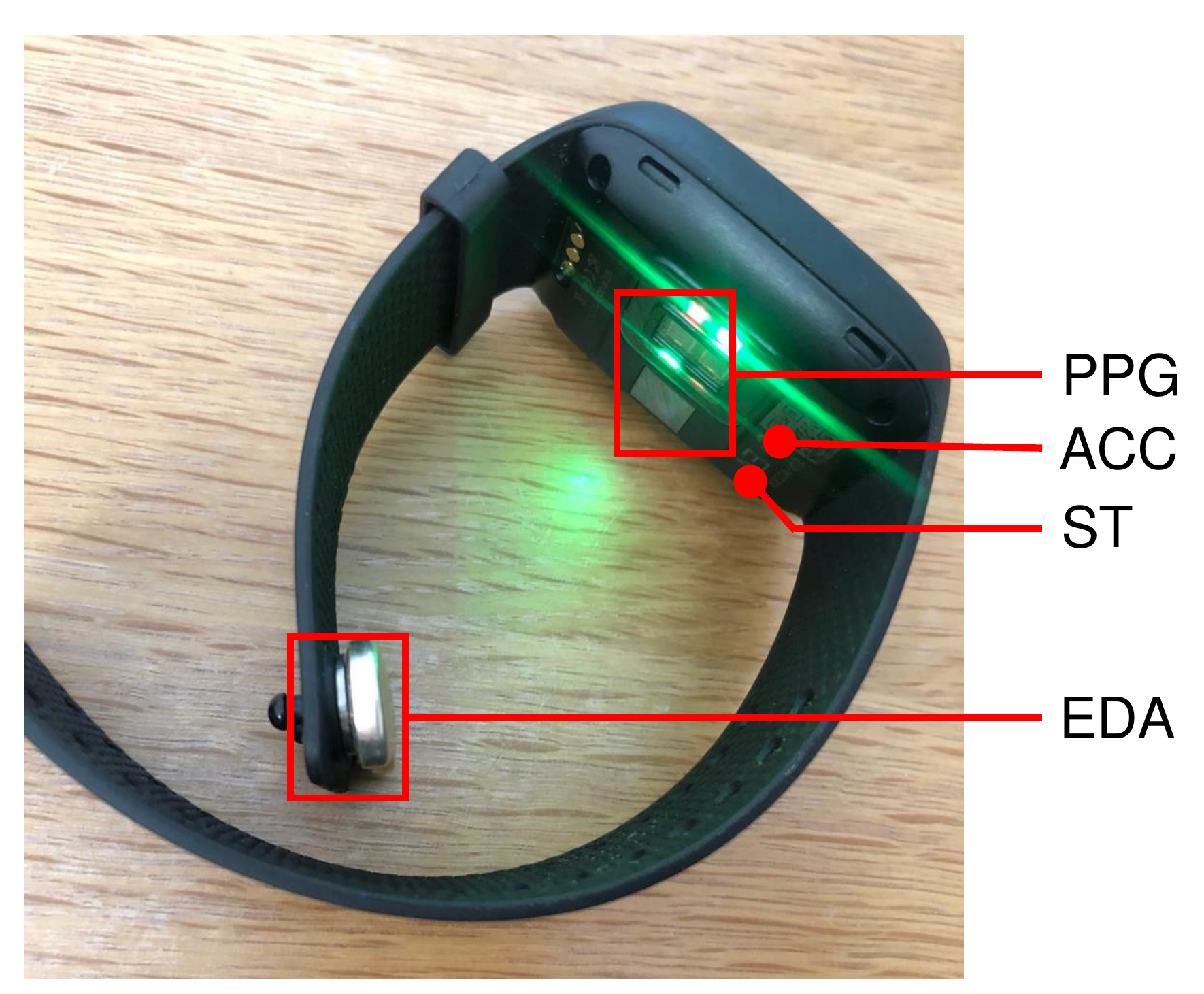}
    \caption{Empatica E4 wristband}
    \label{fig: e4}
\end{figure}

During the data collection, participants were asked to wear \textit{Empatica E4} \footnote{Empatica E4 wristband: \url{https://www.empatica.com/en-int/research/e4/}} wristband as shown in Figure \ref{fig: e4}. E4 wristband is first proposed in \cite{garbarino2014empatica} and has multiple sensors: Electrodermal Activity (\ac{eda}) sensor, 3-axis \ac{acc}, \ac{ppg} sensor, and optical thermometer. \ac{eda} has also been known as galvanic skin response (GSR) or skin conductance, which measures the continuous variation in skin electrical characteristics at 4 Hz. \ac{acc} records the acceleration in three axes at 32 Hz in the range of [-2g, 2g], which captures the physical activity of users. \ac{ppg} is an optically obtained plethysmogram that can be used to measure the Blood Volume Pulse \ac{bvp} at 64 Hz. The \ac{hr} and inter-beat interval (IBI) are derived from \ac{bvp} signals by the wristband. The optical thermometer measures the peripheral skin temperature (\ac{st}) at 4 Hz. Overall, the E4 wristband is light-weight and comfortable, which is particularly suitable for continuous and unobtrusive monitoring of participants in our research. 
It was shown very early on that emotions are related to the autonomic nervous system and that this is accompanied by changes in physiological signals \cite{levenson1988,kreibig2010}. By measuring a person's physiological signals, changes can be recognized, and emotions can be assigned. We use the correlation between mood and physiological signals by extracting features from these signals and incorporating them into our regression models.

\subsubsection{ESM questionnaire}
\label{sec:ESM}
In this study, the participants were asked to rate their mood regarding the last hour (see Figure \ref{fig:mood_esm}). We used the \ac{sam} from Bradley and Lang  \cite{bradley1994measuring} to gather the arousal and valence state. The arousal scale ranges from relaxed to excited, while the valence ranges from positive to negative.
Furthermore, we gathered the dominant social role the person has been in for the last 15 minutes. In \cite{Ashforth2000} the authors describes a social role as a mental construct, which individuals maintain to organize their surroundings. Thus, we investigate \textit{work} and \textit{private} as domains with their labelled social roles as a characterization of different behaviour. Contrary to prior work \cite{NippertEng1996,ozenc2011life}, we decided not to be more granular regarding the different roles, although \textit{family}, \textit{work}, and \textit{social} are reported as the most universal social behaviours. The focus of our study lay on the work-life balance, and the distinction between \textit{social} and \textit{family} seemed redundant, especially since the relationship to contacts covers it.
Finally, we asked the participants for whom they are interruptible - contacts from the \textit{work} or \textit{private} domain, nobody (\textit{none}), or everybody (\textit{both} domains).
\section{Methodology}
\label{sec:method}
\begin{table*}[!t]
	\centering
	\caption{Extracted features divided by device. Information marked with ($\ast$) have been manually reported.}
	\label{table:features}
	\scriptsize
	\begin{tabular}{p{.2\textwidth}p{.5\textwidth}p{.2\textwidth}}
		\toprule                                                  
		\textbf{Feature}
        & \textbf{Description}
        & \textbf{Contextual Information} \\
		\midrule
		
		\multicolumn{3}{l}{\textit{Smartphone Data}}   \\ 
		\cmidrule{1-3}
		topk\_x\_unique
		& Top k applications in the last $x\in{5,10,15,20,25,30}$ minutes.
		& Foreground application \\
		
		phone\_apps\_X
		& Number of used smartphone applications in the last $x\in{5,10,15,20,25,30}$ minutes, extracted from the name and the package identifier of the current foreground application
		& Foreground application \\
		
		
%
%
%
%
        physical\_activity\_X
		& Number of unique physical activities reported by the Google Recognition API
		& Physical activity\\
				place\_top\_x, place\_other
		& Top three ($x\in{1,2,3}$) frequently visited places and all other places. Category of the location according to Google’s Geocoding API.
		& Location (Android)\\
%
		
%
		screen\_on, screen\_off, screen
		& The current state of the screen.
		& Screen state \\

		notification\_length
		& Length of the text within the notification.
		& Notification content \\
		
		Monday, Tuesday, Wednesday, Thursday, Friday, Saturday, Sunday
		& Day of the week.
		& Notification arrival time\\
		
		morning, afternoon, evening, midnight
		& Time of the day: morning (from 6 a.m. to 12 p.m.), afternoon (from 12 p.m. to6 p.m.), evening(from 6 p.m. to 0 a.m.), and midnight (from 0 a.m. to 6 a.m.)
		& Notification arrival time\\
		
		is\_weekend
		& Binary value describing, whether it is weekend or not.
		& Notification arrival time \\
		
		loc\_8, loc\_10
		& Longitude and latitude information of the device as Plus Code
		& Location \\
		
		relation\_x
		& The participant’s relationship to the extracted contact and/or group. Participants could choose between family, friend, work, and none. Multiple selections are possible (e.g., work and friend).
		& Relationship$\ast$\\
		
		contact
		& Hashed contact and/or group name extracted from notification titles
		& Contact$\ast$ \\
		\cmidrule{1-3}
		\multicolumn{3}{l}{\textit{Experience Sampling Method Data}}   \\ 
		\cmidrule{1-3}
		valence, arousal
		&The affective state of the last $60$ minutes
		& Mood$\ast$\\
		
		private, work, both, none
		& Interruptibility preferences of the last $15$ minutes.
		& Interruptibility$\ast$\\
		
		private, work, both
		& Social role of the person in the last $15$ minutes.
		&Social role$\ast$\\
		\cmidrule{1-3}
		\multicolumn{3}{l}{\textit{Physiological Signals}}   \\ 
		\cmidrule{1-3}
		$\mu, \sigma^{2},\sigma$
		& Mean, Variance, Standard Deviation
		& EDA, SCR, SCL, BVP, HR, IBI, ST \\
		min, max
        & Min and max value
        & EDA, SCL, SCR, BVP, HR, ST \\
        rms
        & Root mean square
        & HR \\
		$f_{slope}$
		& The absolute value of the slope of the linear regression line
		& EDA, SCL, HR, ST \\
		$f_{\sqrt{slope}}$
		& The square root of the absolute values of the slope of the linear regression line
		& EDA, SCL, HR, ST \\
		$f_{1 intercept}$
		& The square root of the absolute value of the intercept of the linear regression line
		& EDA, SCL, HR, ST \\
		$f_{2 intercept}$
		& The third power of the square root of the absolute value of the intercept of the linear regression line
		& EDA, SCL, HR, ST \\
		nni\_50/20, pnni\_50/20, nni\_20, pnni\_20
		& Number,and percentage of interval differences of successive RR-intervals greater than 50ms and 20 ms, respectively
		& IBI \\
		vlsf, lf, hf, lf\_hf\_ratio
		& Power in HRV in the very low/low/high frequency. Power of lf/hf
		& IBI \\
		sdsd, range\_nni
		& The standard deviation of differences between adjacent RR-intervals. 
		Difference between the maximum and minimum nn\_interval
		& IBI \\
		cvsd, cvnni
		& Coefficient of variation, of successive differences (cvsd), equal to the ratio of rmssd / sdnn divided by mean\_nni.
		& IBI \\
        triangular\_index
        & The HRV triangular index measurement is the integral of the density distribution divided by the maximum of the density distribution.
        & IBI \\
		\bottomrule
	\end{tabular}
\end{table*}

\subsection{Pre-processing approaches}
In the first part of the machine learning, cleaning up the data is necessary to get rid of noise and homogenize it. This preparation helps to process the data in all further steps.
One of those preparations was to harmonize the applications' names through all considered platforms (windows, macOS, and android), e.g. mapping microsoft-powerpoint to PowerPoint, or by removing system-specific endings by regular expressions.
Furthermore, we opted for parsing the google play store websites according to the mobile applications used by our participants to extract the suited applications category.

The Google Recognition API returns all recognized physical activities and their corresponding confidence ratings. Reducing the data, we chose the activities with the highest confidence rating and forwarded the last known activity for all following events.
An upsampling was also be done for other data like the ringer mode, features regarding the last known locations, and screen status. The software package used to extract more valuable place information was Plus Codes from google \textit{Plus Codes\footnote{See: \url{https://maps.google.com/pluscodes/}}}. The code gives us a description of a rectangular area, including the given longitude and latitude information. Depending on how long the plus code used is, the accuracy of the location information differs.

\subsection{Extracted features}
We prepared the data according to our needs for the regression model. It was decided that the best method for this investigation was to calculate the features on the data before the notification arrived. 
All extracted features are shown in Table ~\ref{table:features}.

\subsubsection{Features extracted from Smartphone Data}

We first examined the current context of the user. On the one hand, it can be deduced from this whether the user is currently interruptible and, accordingly, whether the user will react immediately to an incoming notification or not. For this purpose, we analyzed the apps used in the last 5 to 30 minutes until the notification arrived.
We discovered the top \hl{\textit{k}} smartphone applications by counting the appearance of the application per user. Assume user $X_1$ has an app set $\mathcal{A}=\{A_1,A_2,…,A_N\}$, where the app is sorted according to the number of receiving notifications from the training dataset. \hl{Namely}, $A_1$ app receives most notifications and $A_n$ receives least notifications. In this research, we only study the top-k apps  $\mathcal{A}= \{A_1,A_2,…,A_k\}$, where the $k$ is set to be \hl{10}. \hl{We will explain the $k$ in detail in Section} \ref{sec: individual diff}.


One indicator of whether the person would respond immediately to a notification is whether the smartphone is currently in use. For this, we asked whether the screen was on or not. We also took into account the length of the notification and from whom the message came from. If the contact was known, we included the relationship to this contact as well. The sender-recipient relationship is closely related to the response rate of the notification \cite{Mehrotra2015,Mehrotra2016}. Mehrotra et al. have reported that the sensed interruption depends on the sender of a notification and that chat notifications from a family member or relative have shown the highest acceptance rates.
As described earlier, we used the Android Google API to record the current physical activities of the participants. Breakpoints in physical activities have been proven to mark opportune moments for interruptions. Okoshi et al. \cite{Okoshi2015,Okoshi2015a,Okoshi2017} examined breakpoints within physical activities and application usage. The authors find that notifications delivered at breakpoints denoted as transitions between applications and physical activities can lower the individuals' mental burden. Ho and Intille \cite{Ho2005} also suggest that notifications delivered during activity transitions produce more favorable outcomes than those delivered randomly. The number of different activities detected was also used as a feature in the first stage classification. As another feature, we used the Plus Codes representing the location where the user is currently staying. The most frequently visited locations of each subject were also set as features. For this purpose, we first determined the three locations of each subject visited most frequently by him during the measurement process. These three locations represent our top 3 locations. All other Plus Codes were assigned to the category other. After that, it was determined where the participant has been before receiving the notification by setting one of the top 3 locations, or the category other true.

Furthermore, we used the day of the week, whether it was a weekend, and the time of day as features representing time. To set a time of the day, we split the day into four parts, midnight (from 0 a.m. to 6 a.m.), morning (from 6 a.m. to 12 p.m.), afternoon (from 12 p.m. to 6 p.m.), and evening (from 6 p.m. to 0 a.m.). Several previous studies have investigated the relationship between times of the day and notification response \cite{Okoshi2017,Okoshi2019,Saikia2017}. Okoshi et al. and Saikia et al. have found that by sending out notifications at opportune times of the day, the response time is greatly decreased.



\subsubsection{Features from ESM Data}
Additionally to the other features, we used the ESM questionnaire data, describing the mood, the interruptibility preferences and the current social role. The mood was measured in two scales: valence and arousal (see Section \ref{sec:ESM}). They represent different kinds of feelings: from unhappy to happy (represented as 1-5) and from calm to excited (represented as 1-5), respectively.
In Section \ref{sec:EDA} we used the features containing the contextual information of interruptibility and social role. We applied one-hot-encoding to represent this nominal data.

\subsubsection{Features for Physiological Signals}

We decided to extract statistical features on all given physiological signals, which are common to be used for mood recognition. Furthermore, we followed Heinisch et al. \cite{heinisch2019}, adding features based on the linear regression line. These features have been shown to be robust to the influence factor of physical activity. Since we also conducted an in-the-wild study, we fell back on this type of feature.

The EDA signal can be divided into two components, the skin conductance response (SCR) and the skin conductance level (SCL). SCR contains the high-frequency components of the signal, reflecting the rapid changes in the signal in response to a stimulus. In contrast stand SCL, which contains the low-frequency components of the EDA  and thus represents the long-term or baseline conductance. Splitting the EDA signal into these two components, the python tool of Greco et al. \cite{greco2016} was used.

\begin{table}[b]
\caption{{Notification} and \hl{app} information for 18 participants}
\label{tab:overview}
\small
\begin{tabular}{@{}lllll@{}}
\toprule
                                                                                           & Min     & Max     & Median  & Mean    \\ \midrule
Number of apps                                                                             & 18      & 47      & 26     & 30      \\ \hline
Number of notifications                                                                    & 363     & 6213    & 1914    & 2362    \\ \hline
\begin{tabular}[c]{@{}l@{}}Percentage of notifications\\ sent by top 10 apps\end{tabular} & 84.67\% & 99.57\% & 94.88\% & 94.30\% \\ \hline
\begin{tabular}[c]{@{}l@{}}Percentage of notifications\\ sent by top 5 apps\end{tabular} & 65.56\% & 97.90\% & 84.17\% & 83.33\% \\ 
\bottomrule
\end{tabular}
\end{table}
\section{Understanding the Mood, Usage Behaviours and Notification Response Time of Participants}
\label{sec:EDA}

\hl{In total, we have received 3236 ESM responses from {$18$} participants during the data collection. First, we explore the notification response patterns of different participants. Second, we investigate the relationship between users' mood and their notification response time. Finally, we explore how mobile usage behaviours are related to notification response time. }

\subsection{Understanding Notification Response Times for Different Participants}
\label{sec: individual diff}

\begin{figure}
    \centering
    \includegraphics[width=0.55\textwidth]{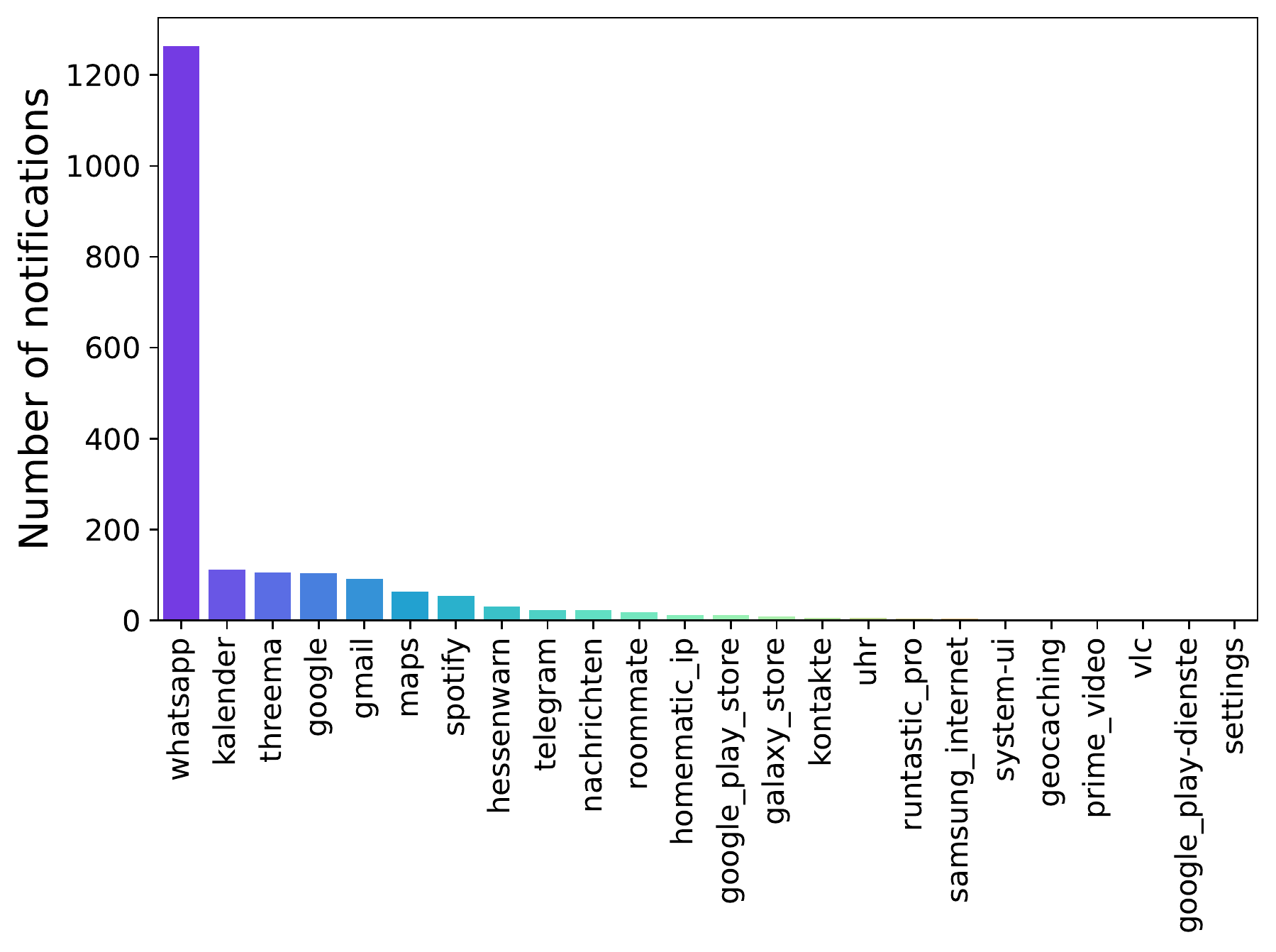}
    \caption{The number of notifications across all the apps for participant P10}
    \label{fig:dis_no_par}
\end{figure}

We explore the notification response time from top-$k$ apps where $k$ = 10 because on average, the top ten apps sent 94.30\% {of the} notifications (out of 2,362 notifications), while {the other apps} only sent 5.70\% {of the} notifications (see Table \ref{tab:overview}). If we only study the top five apps, we would miss 16.67\% {of the} notifications, which is almost three times the number of missed notifications {from} studying the top ten apps. \hl{For instance, Figure }\ref{fig:dis_no_par} \hl{displays the number of notifications across all the apps for one participant P10 during the data collection. We find that P10 received 96.04\% notifications from top ten apps and 86.16\% from top five apps.} Therefore, in this research, we {did} not consider the apps receiving {only a few} notifications ($k$ > 10) because {the} relatively small dataset \hl{would not offer a robust representation of the notification response times for} modelling. In real-world scenarios, $k$ can be set to {any} values based on the categories of apps {being} explored.

\begin{figure}
	    \centering  
		\includegraphics[width=0.65\textwidth]{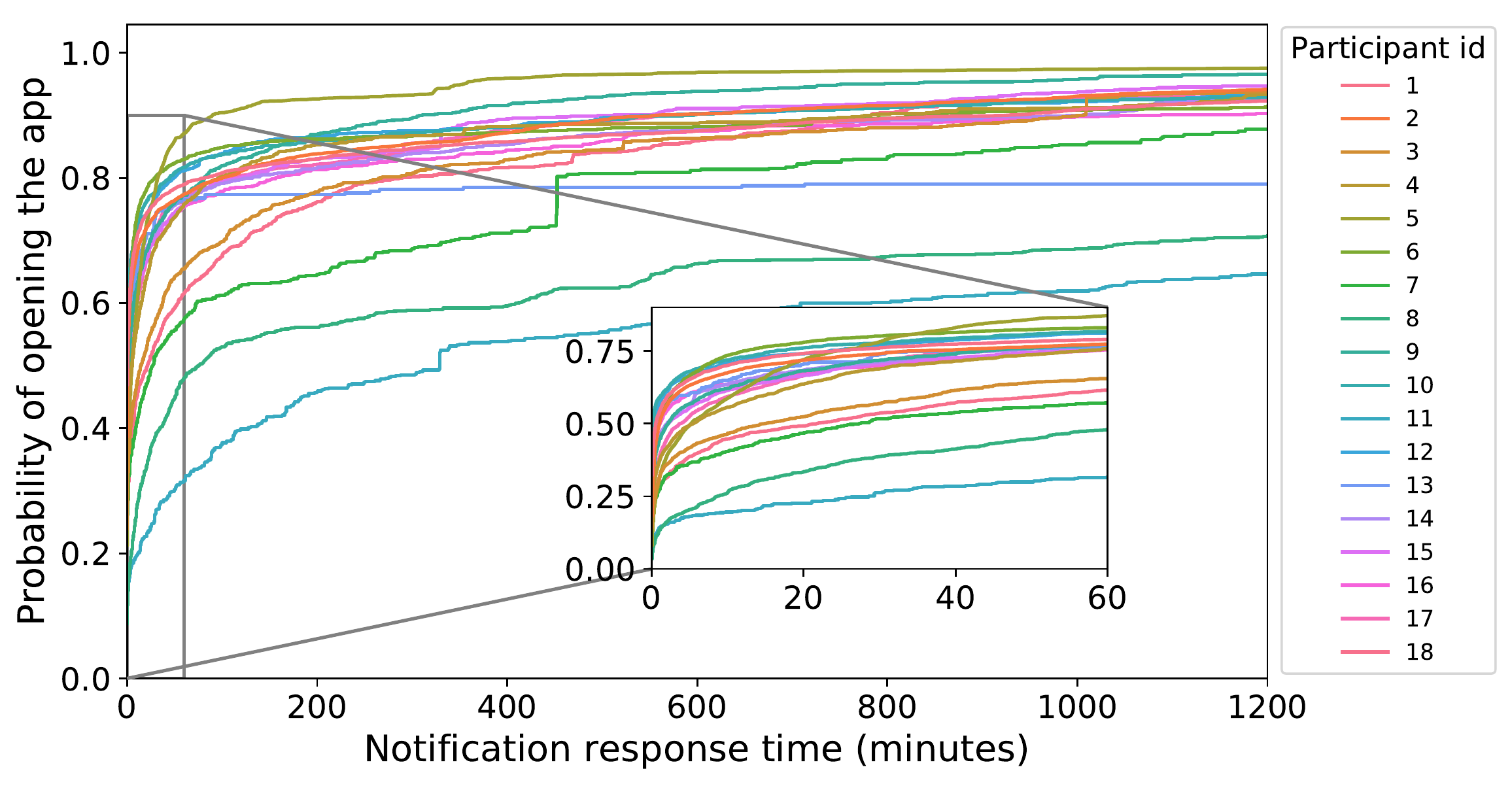}
		\caption{
Cumulative distribution of notification response time from top ten apps for all participants}
		\label{fig:cdf}
\end{figure}

To understand the notification response time for all participants,  we show the cumulative distribution of notification response times from top ten apps for each participant in Figure \ref{fig:cdf}. It is obvious that response time to most notifications is short, but the response time of some notifications are long. Specifically, out of 40,290 notifications received by 18 participants, the response time was within five minutes for 54.32\% of the notifications, within one hour for 75.86\% of the notifications, within one day for 93.90\% of the notifications. However, if we look at the response times for different participants, we find that each participant has their own patterns and trends for responding to  notifications. For instance, participant P5 responded to 49.37\% of their notifications within five minutes and 86.96\% within one hour, while participant P11 responded to notifications much more slowly, only responding to 18.46\% within five minutes and 32.36\% within one hour. Hence, studying the participant-wise notification response time is necessary, as the general model may be inaccurate due to individual differences.




\begin{figure}	
    \begin{subfigure}{0.49\textwidth}
	    \centering
		\includegraphics[scale=.49]{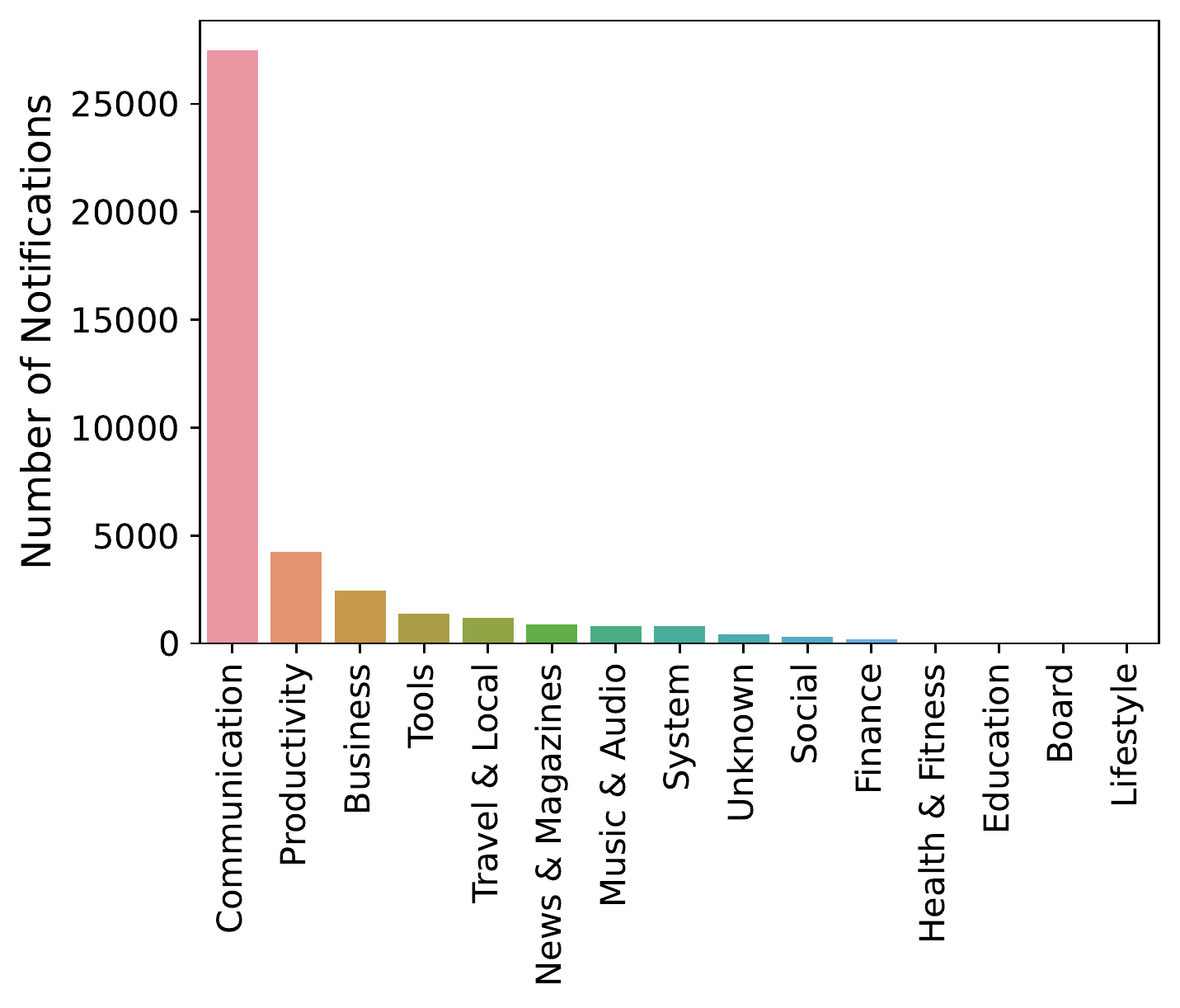}
		\caption{Number of notifications}
		\label{fig:app_cat_num}
	\end{subfigure}
	\begin{subfigure}{0.49\textwidth}
	    \centering  
		\includegraphics[scale=.49]{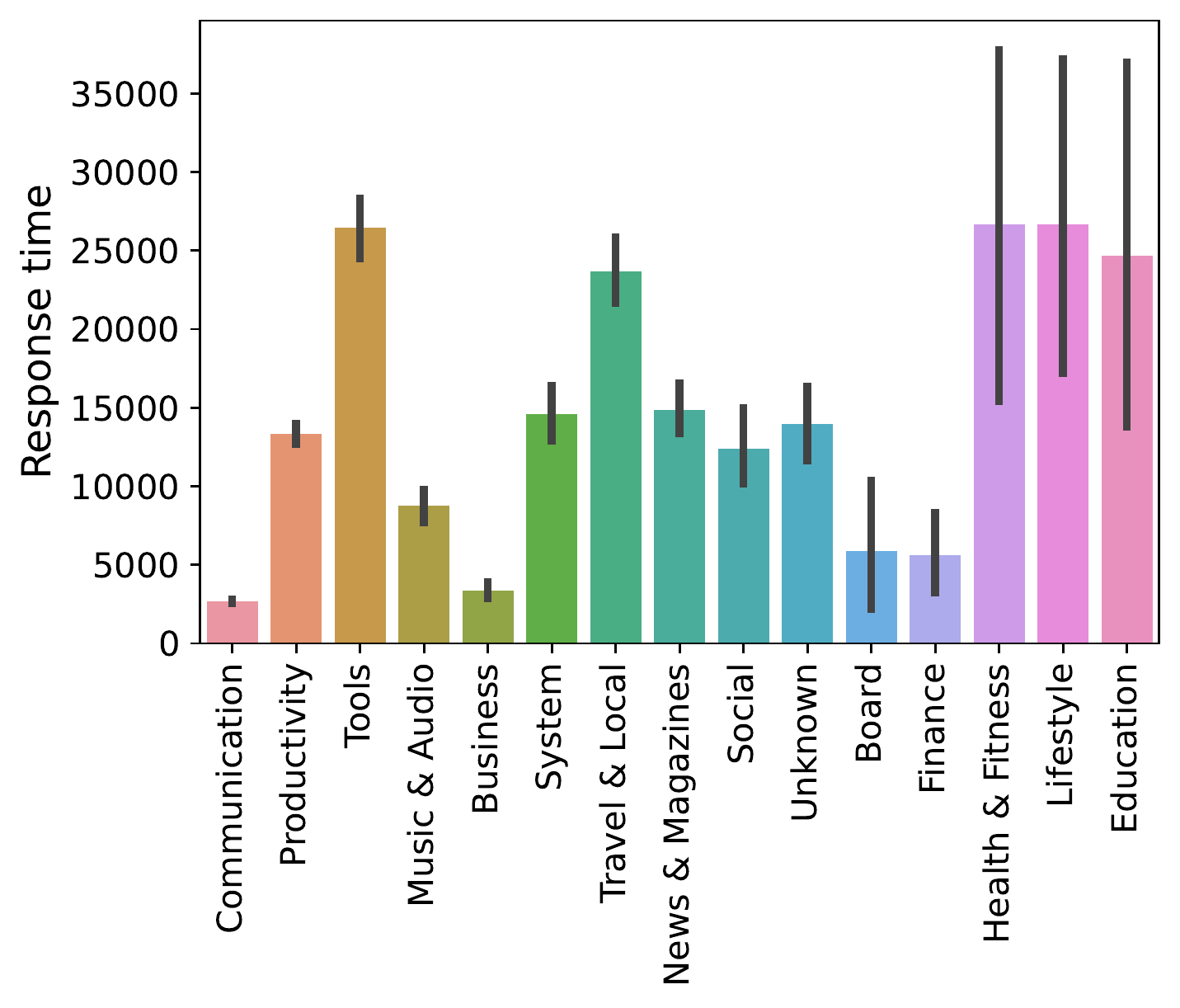}
		\caption{Average response time}
		\label{fig:app_cat_time}
	\end{subfigure}
	\caption{Information for different app categories}
	\label{fig:app_cat}
\end{figure}

\subsection{App Categories and Response Time}
Figure \ref{fig:app_cat_num} displays the number of notifications across the app categories, showing that the \textit{communication} apps receive much more notifications than all the other app categories. In total, \textit{communication} apps receives six times more notifications than the app category that was ranked second (i.e. \textit{Productivity} apps).  Figure ~\ref{fig:app_cat_time} shows the average response times for each app category (black vertical line indicates the error bar, with a 95\% confidence interval). \hl{Since  93.90\% of the notifications from all participants are responded in one day, we focus on analysing those notifications and have removed the notifications with a response time of more than one day. Messages that have not been responded more than 24 hours may be due to various reasons, such as the user forgot, or has already responded on other platforms. We believe that it is more meaningful to focus on the notifications that users reply in a timely manner, and the small number of notifications unanwsered for a long time will be explored in our future research.} We find that the response times varied significantly between the app categories. If we aim to predict response time across all categories, the prediction performance would be unreliable due to the extreme variations in the number of notifications and the average notification response time between app categories. Therefore, in this research, we focus on predicting users' response behaviours for \textit{communication} apps. 

\subsection{The Mood of Users and Notification Response time}

\begin{figure}
\begin{subfigure}{0.4\textwidth}
	    \centering
		\includegraphics[scale=.36]{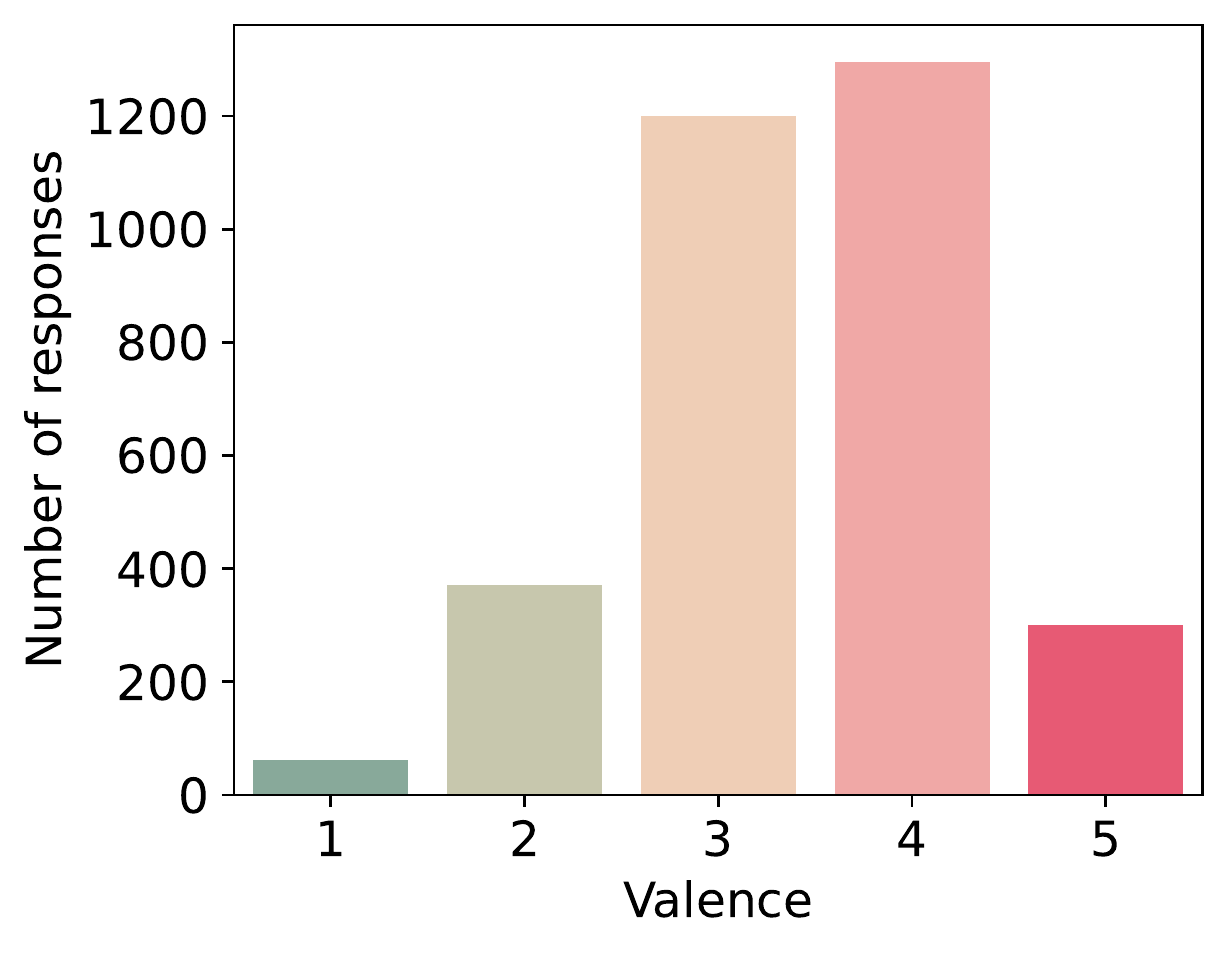}
	\end{subfigure}
\begin{subfigure}{0.4\textwidth}
	    \centering  
		\includegraphics[scale=.36]{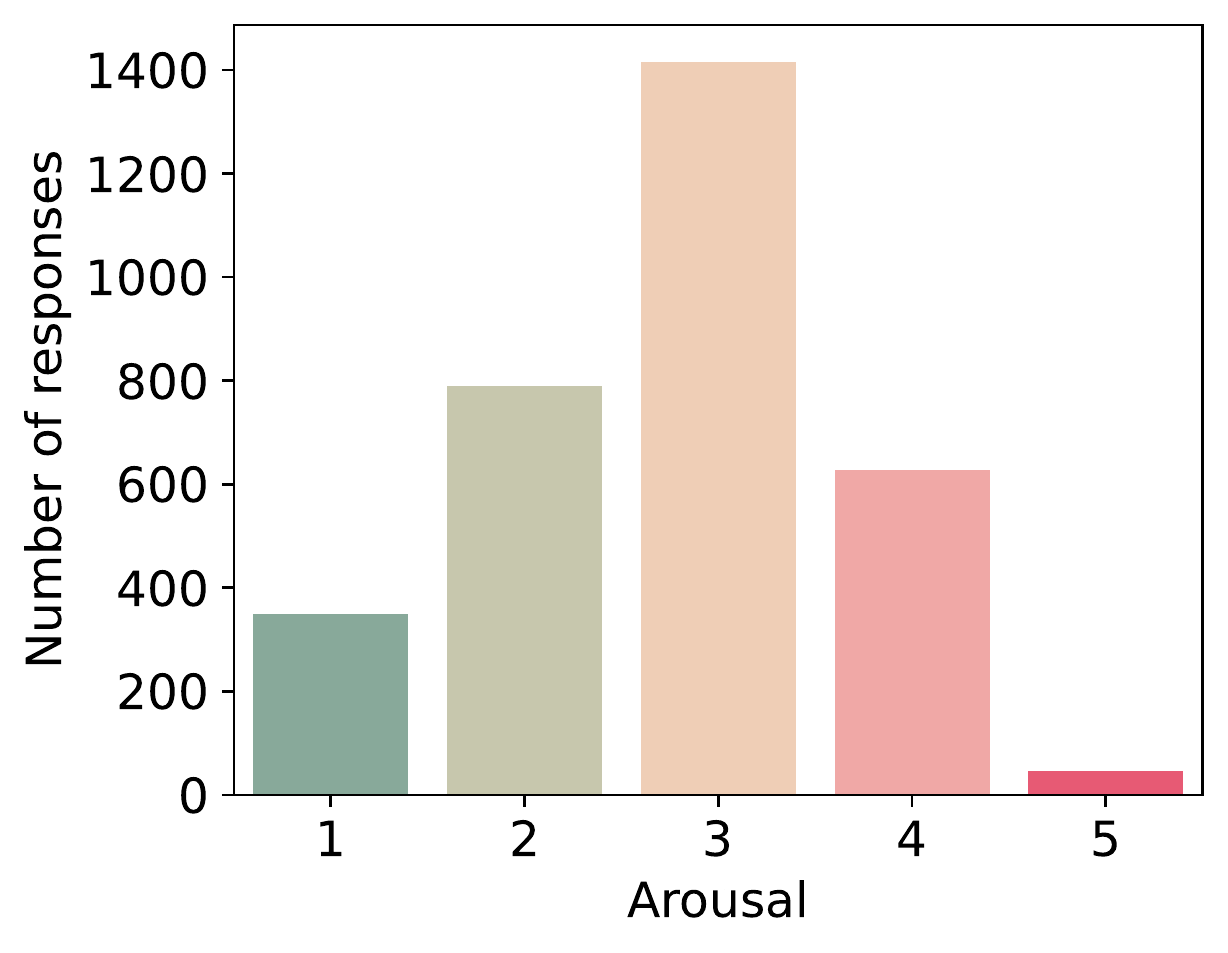}
		
	\end{subfigure}
	\caption{Distribution of arousal and valence for 18 participants}
	\label{fig:dis_valence_arousal}
\end{figure}

\begin{figure}[b]
	\begin{subfigure}{0.45\textwidth}
	    \centering
        \includegraphics[scale=.5]{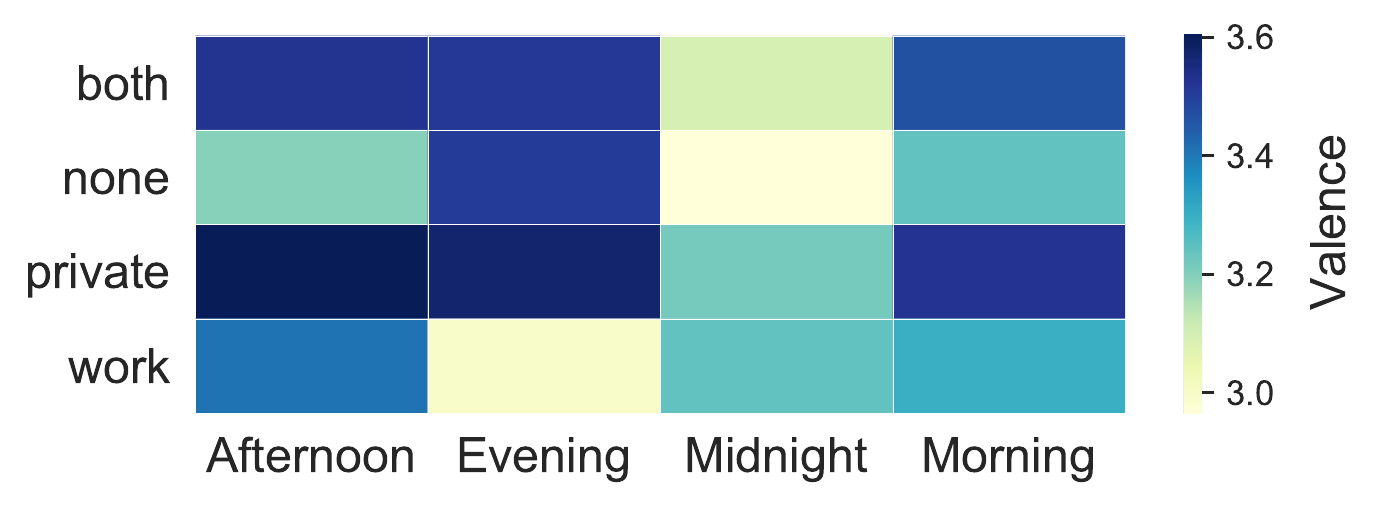}
		\caption{Valence}
		\label{fig:valence_time}
	\end{subfigure}
	\begin{subfigure}{0.45\textwidth}
	    \centering  
		\includegraphics[scale=.5]{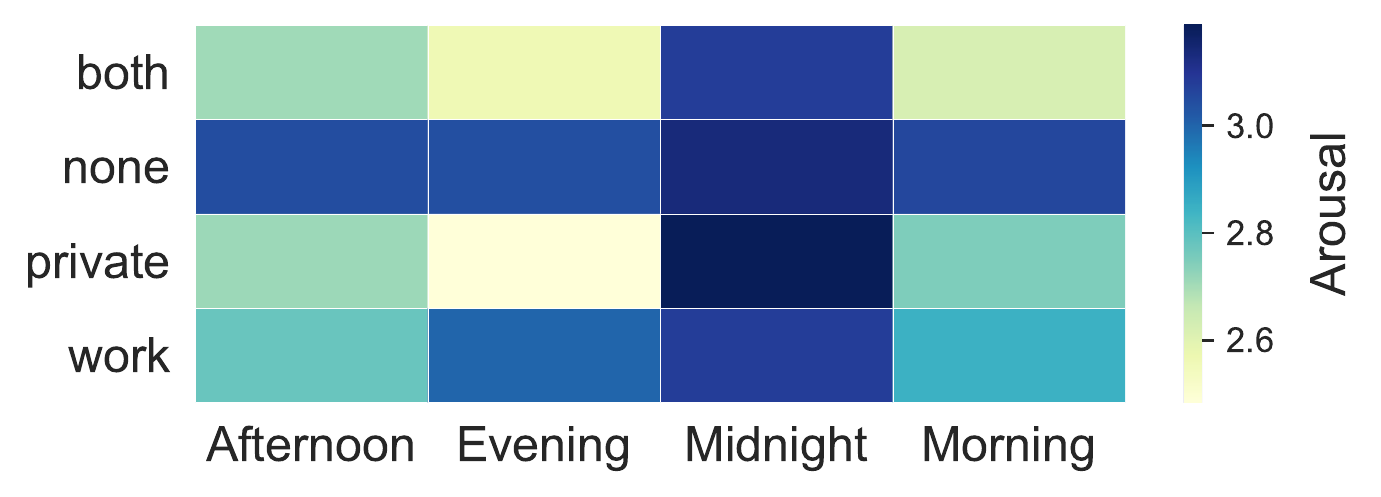}
		\caption{Arousal}
		\label{fig:arousal_time}
	\end{subfigure}
	\caption{{Mood} of participants {at} different {levels of} interruptibility and {various times of the day}}
	\label{fig:mood_interruptibility_time}
\end{figure}

We calculate \hl{the overall} distribution of mood in Figure \ref{fig:dis_valence_arousal}, where \hl{1 to 5} indicates \hl{a} low to high value of valence/arousal. Generally, participants usually \hl{reported} positive valence {(mean = 3.44)} and low arousal \hl{(mean = 2.77), meaning that they were} relaxed, clam, \hl{and comfortable} \cite{russell1980circumplex} \hl{most of the} time. 
We also explore how the mood \hl{is related to factors such as daytime and} interruptibility. 
As shown in Figure~\ref{fig:valence_time} and Figure~\ref{fig:arousal_time}, participants usually {experienced} the highest valence (mean = 3.56) {and} lowest arousal (mean = 2.58) in the evening (6pm - 12am). In contrast, they usually experienced the lowest valence (mean = 3.28) {and} highest arousal (mean = 2.97) in the midnight (12am{-}6am). We also {found} that when {the} participants {did not} want to be interrupted by {either} work or private affairs ({i.e. interruptibility was 'none'}), they {were} usually {experiencing} lowest valence {(mean = 3.21)} and highest arousal{(mean = 3.03)}. Interestingly, when the participants \hl{experienced positive mood} (high valence), they \hl{were} more likely to be \hl{amenable to interruptions relating to} private{, or private and work (i.e. both)} affairs. \hl{In general}, \hl{the} participants \hl{experienced} \hl{varying mood} with different \hl{levels of} interruptiblity \hl{at different times}.

\begin{figure}[]
	\begin{subfigure}{0.45\textwidth}
	    \centering
        \includegraphics[scale=.5]{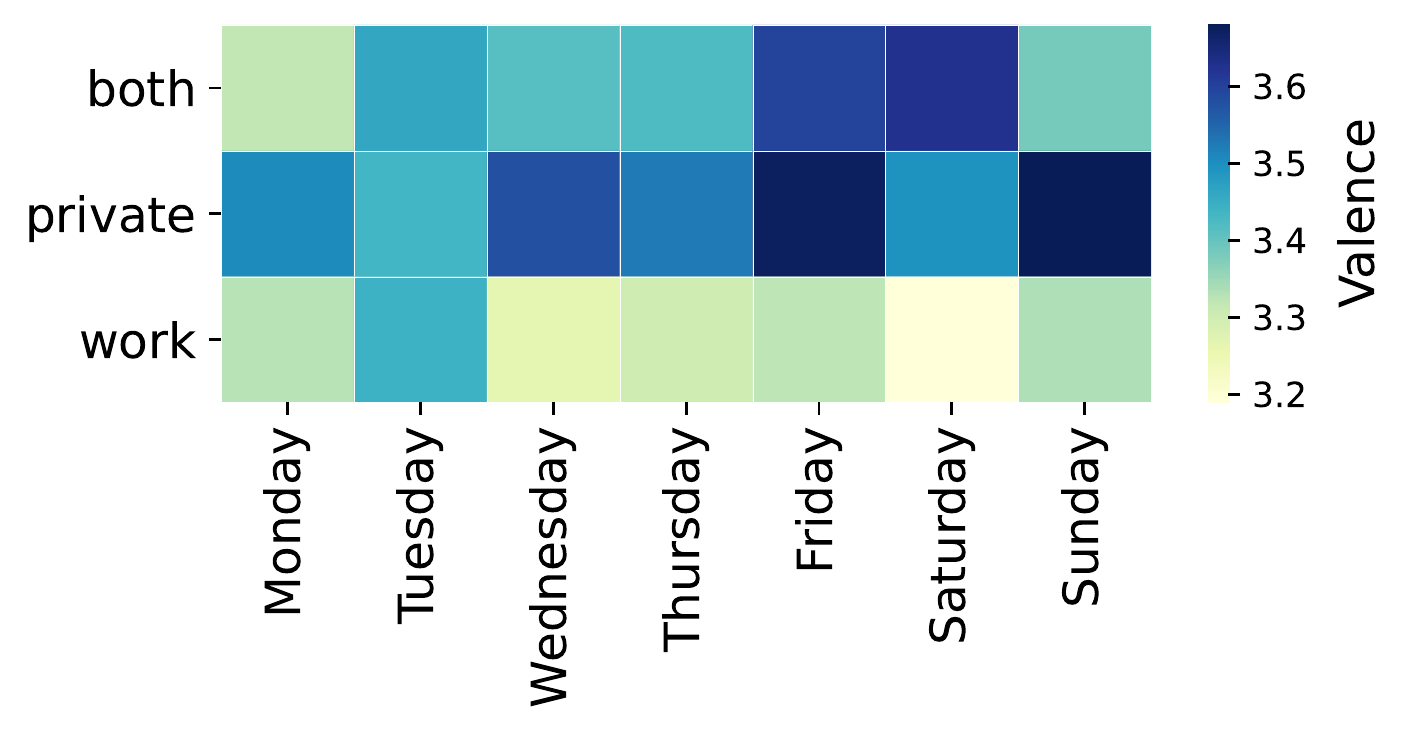}
		\caption{Valence}
		\label{fig:valence_weekday}
	\end{subfigure}
	\begin{subfigure}{0.45\textwidth}
	    \centering  
		\includegraphics[scale=.5]{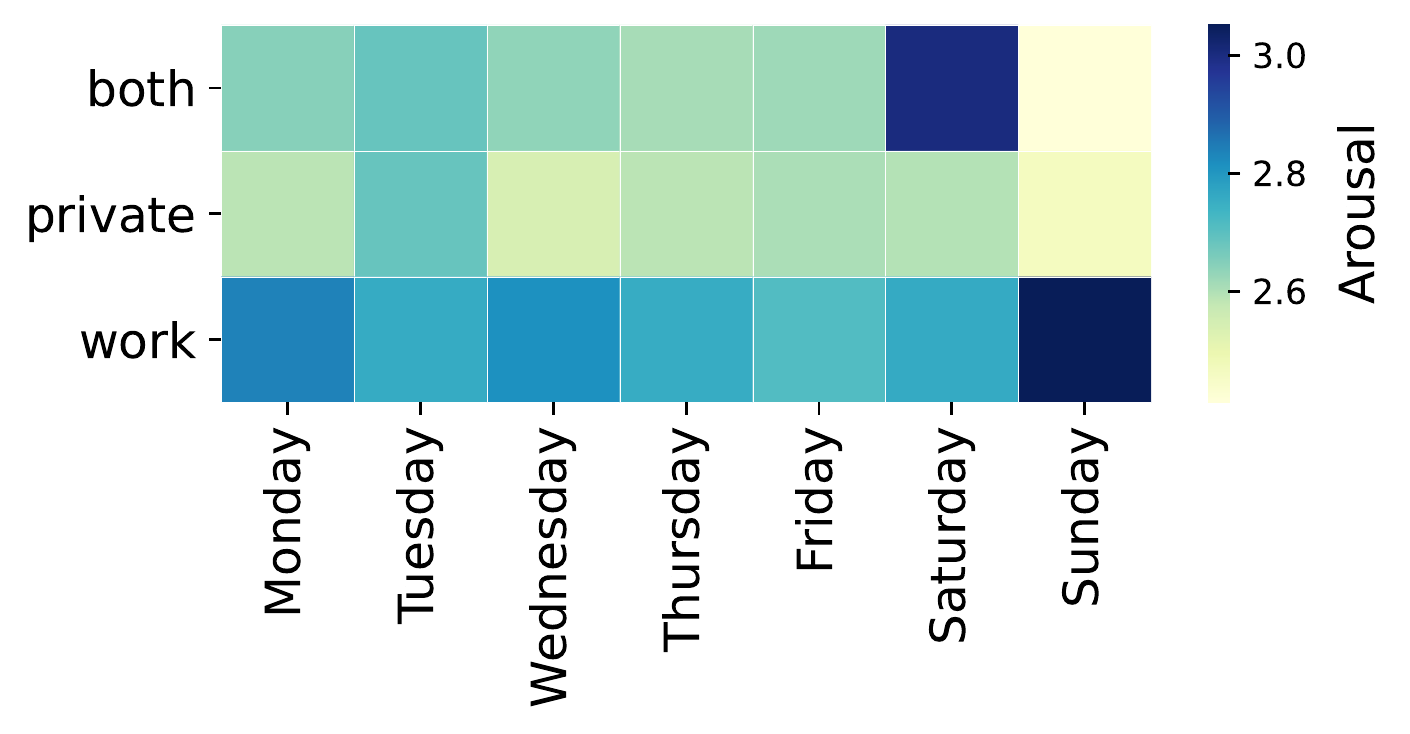}
		\caption{Arousal}
		\label{fig:arousal_weekday}
	\end{subfigure}
	\caption{{Mood} of participants over different social roles and {days of the week}}
\end{figure}

We also investigate how the mood {changed based on} social roles and {the day of the week} (see Figure \ref{fig:valence_weekday} and Figure \ref{fig:arousal_weekday}). We {found} that participants usually {experienced} high valence {(mean = 3.55)} and low arousal {(mean = 2.67)} when they {were} busy with private issues {and tended} to {experience} low valence {(mean = 3.31)} and high arousal {(mean = 2.88)} when they {were at} work. \hl{Our participants had the highest valence in the private role on Friday (mean=3.66) and Sunday (mean=3.66) and the lowest valence values (mean=3.25) at work on Saturday. Saturday and Sunday were also different in the arousal scale, as the social roles 'both' (mean=3.08) and 'work' (mean=3.22) had the highest values, respectively. Interestingly enough, being in the role of private or both made our participants feel the lowest arousal (mean=2.5) on Sunday.}

Mehrotra et al. \cite{mehrotra2017mytraces} investigated the causal links between users' {emotions} and their {interactions} with mobile phones. They found users’ emotions {had} a causal impact on mobile phone {interactions}. In this research, we {investigated} the relationship between the {participants} arousal/valence and {their} notification response {times} to understand whether {human affect (in our case mood)} can be used as {a} proxy {for user} response behaviours. We {computed} the \textit{Spearman Rank Correlation} \cite{de2016comparing} because the results from {the} \textit{Shapiro-Wilk Test} \cite{razali2011power} and \textit{D'Agostino's Test} \cite{d1971omnibus} \hl{revealed that none of the samples were normally distributed} (p $\leq$ 0.05). We  {found} that both the valence and arousal  {were} significantly correlated with the notification response time: Valence was negatively correlated (r = -0.04, p $\leq$  0.001) and arousal  {was} positively correlated (r = 0.02, p = 0.012).  {These} results indicate that people usually  {took} longer time to respond to notifications when they are distressed, frustrated or angry (low valence and high arousal). Therefore, we will take into account the influence of valence and arousal in modelling the notification response  {times} in Section \ref{sec:experiment}.



\begin{figure}[b]
	    \centering  
		\includegraphics[width=1\textwidth]{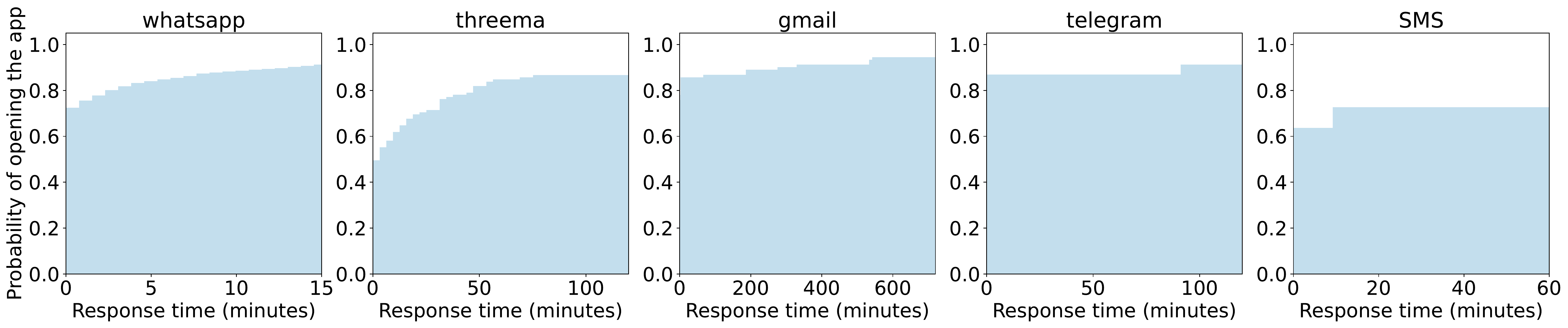}
		\caption{
Cumulative distribution of notification response times from five apps for participant 10}
		\label{fig:cdf_9apps}
\end{figure}

\subsection{Impact of Applications on Notification Response Times}

We already know that each participant has their own patterns for responding to notifications. However, we also investigate whether each participant respond to different apps in different ways. Here we explore the influence of apps on  notification response times. Figure ~\ref{fig:cdf_9apps} shows the cumulative distribution of the notification response times for five popular apps for participant P10. It clearly shows that even for the same participant, the notification response times vary from app to app. For example, this participant usually responded quickly to \textit{whatsapp}, \textit{gmail} and \textit{telegram} but much more slowly to \textit{threema}. Specifically, within five minutes, this participant responded to 83.53\% of notifications from \textit{whatsapp} but 53.33\% from \textit{threema}. 
Therefore, it is necessary to consider the impact of the apps to meaningfully model the notification response times.



\section{Experiment}
\label{sec:experiment}
As introduced in Section \ref{sec:participants_intro}, we will only focus on predicting the response times of 18 participants who installed the smartphone app. In this research, we built the regression model for predicting the users' response time to notifications. Firstly, we introduce the experiment settings and prediction pipeline. Then we show the overall prediction results and study the impact of mood-related features. Lastly, we investigate how individual differences or categories of applications influence the response time.

\subsection{Prediction Pipeline}
We adopted the regression model for predicting notification response times. The prediction pipeline is described below. 



\textbf{Regressors}.
In the prediction model, we adopted 
several commonly used regressors such as \textit{Standard Linear Regressor} \cite{seber2012linear}, \textit{Support Vector Regressor} (SVR) \cite{cawley2004fast}, \textit{Gradient Booting Regressor} (GBR), \textit{Randome Forest Regressor} \cite{liaw2002classification} and \textit{Bayesian Ridge Regressor} \cite{tipping2001sparse}. Linear regressor is one of the most widely used regression models. The \textit{Support Vector Machine} (SVM) in regression problems is usually known as SVR, which is one of the most commonly used regression models. The GBR model is a powerful prediction model, and it is an ensemble method combining a set of weak predictors to achieve reliable and accurate predictions. \textit{Random Forest Regressor} follows the idea of the random forest, and it can estimate the importance of various features in a model. \textit{Bayesian Ridge Regressor} conducts linear regression using probability distributors rather than point estimates, which provides a natural mechanism to create predictive models when data is insufficient or poorly distributed.


\textbf{Validation}.
Cross-validation is a common practice for training and testing prediction models and is used to estimate the unbiased generalisation performance of models. However, cross-validation may lead to the optimistically biased evaluation of prediction performance when the same cross-validation process is chosen to both tune and select the model. Similar to previous ubiquitous computational studies \cite{gao2020n,di2018unobtrusive}, we adopted \textit{nested cross-validation} \cite{muller2016introduction}, which performs two iterations over the data. The outer loop is used to evaluate the performance of the regressors, and the inner loop is used for optimisation of hyper-parameters and feature selection. After performing this cross-validation, we then applied
\textit{k-fold cross-validation} (\textit{k} = 5) on both loops for each participant. In the outer loop, once the training set and testing set were defined, we standardised features by removing the mean and scaling the data to unit variance. In the inner loop, we optimised the hyper-parameters using a grid search. We then selected features according to the \textit{K} highest scores based on \textit{f-regression} \cite{pedregosa2011scikit} (f-value between the label/feature for regression tasks). \hl{The top eight features} were selected as the input features for each regression model because we found that the this resulted in the lowest prediction error.

\begin{figure}[]
	\begin{subfigure}{0.95\textwidth}
	    \centering
        \includegraphics[width=1\textwidth]{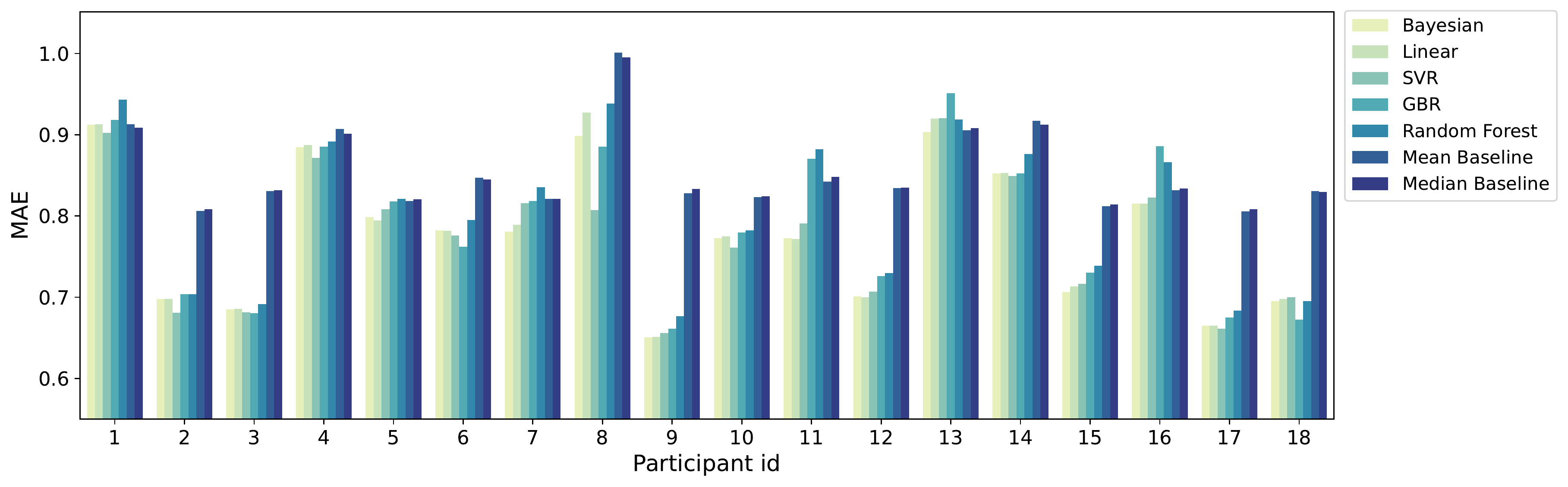}
		\caption{MAE result}
		\label{fig:mae_withoute4}
	\end{subfigure}
	\begin{subfigure}{0.95\textwidth}
	    \centering  
		\includegraphics[width=1\textwidth]{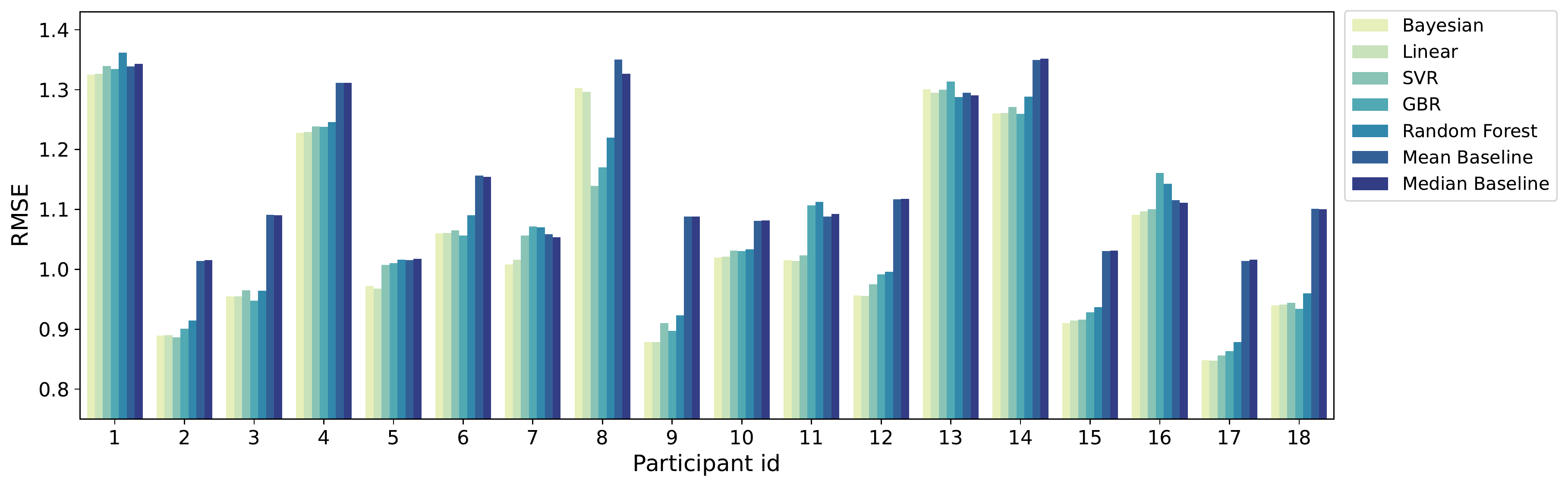}
		\caption{RMSE result}
		\label{fig:rmse_withoure4}
	\end{subfigure}
	\caption{Prediction results across different regressors for each participant}
	\label{fig:prediction_individual}
\end{figure}
\textbf{Baselines}.
\hl{In human-centred research, it is usually difficult to compare the prediction results with state-of-art baselines. The main reason is that the types of data collected, the demographics of participants and the natural environment vary widely across studies, it is not fair or applicable to compare the prediction performance between different studies. Additionally, to our knowledge, we have not found any research that attempts to predict the notification response time for mobile users. As a result, similar to previous human-centred 
studies} \cite{gao2020n, wang2018sensing}, \hl{we have adopted simple baselines to compare the modelling performance. In particular, we} compare the proposed models with two baselines: \textit{Mean} baseline and \textit{Median} baseline. \hl{As one of the most widely used simple baselines to compare with other regressors,} \textit{Mean} baseline always predicts the mean of the training set. \textit{Median} baseline always predicts the median of the training set. \hl{The reason why we choose Median baseline is that the distribution of notification response time is highly skewed (see Figure} \ref{fig:cdf}), \hl{whereas the Median baseline is most informative for skewed distributions or distributions with outliers.}

\hl{}\textbf{Evaluation Metrics}.
To evaluate the performance of \hl{notification response time}, \textit{the Mean Absolute Error} (MAE) and \textit{Root Mean Squared Error} (RMSE)  metrics are applied for evaluating the prediction performance. \hl{The MAE = $ \frac{1}{n} \sum_{i=1}^{n} |y_{true}-y_{pred}|$ and RMSE = $\frac{1}{n} \sum_{i=1}^{n} (y_{true}-y_{pred})^2$, where $n$ indicates the number of samples, $y_{true}$ means the actual notification response time and $y_{pred}$ means the predicted response time. The MAE and RMSE describe the goodness of predictions compared with the ground truth of notification response time. The closer the MAE and RMSE are to 0, the better the performance of the prediction model.}

\subsection{Prediction Result with Mobile Data}

As discussed in Section \ref{sec: individual diff},  the notification response behaviours were very different between the participants (see Figure~\ref{fig:cdf}). Therefore, in the experiment, we built participant-wise regression models instead of a general model for all participants. Figure~\ref{fig:mae_withoute4} and Figure~\ref{fig:rmse_withoure4} show the MAE and RMSE results across different regressors for each participant. We found that the regression models achieved much better predictive performance than both baselines for most participants (i.e., P2, P3, P9, P12, P17 and P18). For example, for participant P9, the \textit{Bayesian} regression model had the best predictive performance (MAE = 0.6505 and RMSE = 0.8779), with  0.1828 (21.94\%) of MAE and 0.2101 (19.31\%) of RMSE lower than the \textit{Median} baseline model.

However, for some particular participants (e.g., P1 and P13), only a small number of regressors achieved lower MAE and RMSE than the baseline models. The possible reasons why some regressors did not work well on a small number of participants are twofold: (1) The notification response behaviours of these participants were more random and changeable than others, which makes them difficult to predict. These individual differences in mobile usage behaviours have been discussed in prior research \cite{cassitas2019study}. (2) These participants had very different notification response behaviours when using different apps, which is difficult to represent in one regression model. However, it was not practical to build a predictive model for each app due to the limited number of notifications.

Next, we calculated the overall prediction performance for all participants by averaging the MAE and RMSE values from the participant-wise models. Table ~\ref{tab:overallresult} shows the overall prediction result for all participants. It shows that all regression models had better prediction performance than the two baseline models in terms of MAE and RMSE, demonstrating the models' potential for predicting notification response time for the ordinary people. The \textit{Bayesian} model achieved the best prediction performance of all the regression models and obtained the 0.7764 of MAE and 1.0527 of RMSE, which was 0.078 (9.10\%) and 0.093 (8.09\%) lower than the mean baseline, respectively. Although the overall prediction performance does not sounds particularly good, the prediction performance was very high for most individuals (see Figure~\ref{fig:prediction_individual}). 

\begin{table}
\caption{Prediction results with different regressors using mobile data}
\begin{tabular}{@{}llllllll@{}}
\toprule
     & \textit{Bayesian.}    & \textit{Linear.} & \textit{SVR.} & \textit{GBR.} & \textit{R. Forest.} & \textit{Mean Baseline} & \textit{Median Baseline} \\ \midrule
MAE  & {\ul \textbf{0.7764}} & 0.7797           & 0.7770        & 0.7797        & 0.8014                  & 0.8541                 & 0.8544                   \\
RMSE & {\ul \textbf{1.0527}} & 1.0533           & 1.0601        & 1.066         & 1.0798                  & 1.1454                 & 1.1441                   \\ \bottomrule
\end{tabular}
\label{tab:overallresult}
\end{table}

\begin{figure}
    \centering
    \includegraphics[width = 0.8\textwidth]{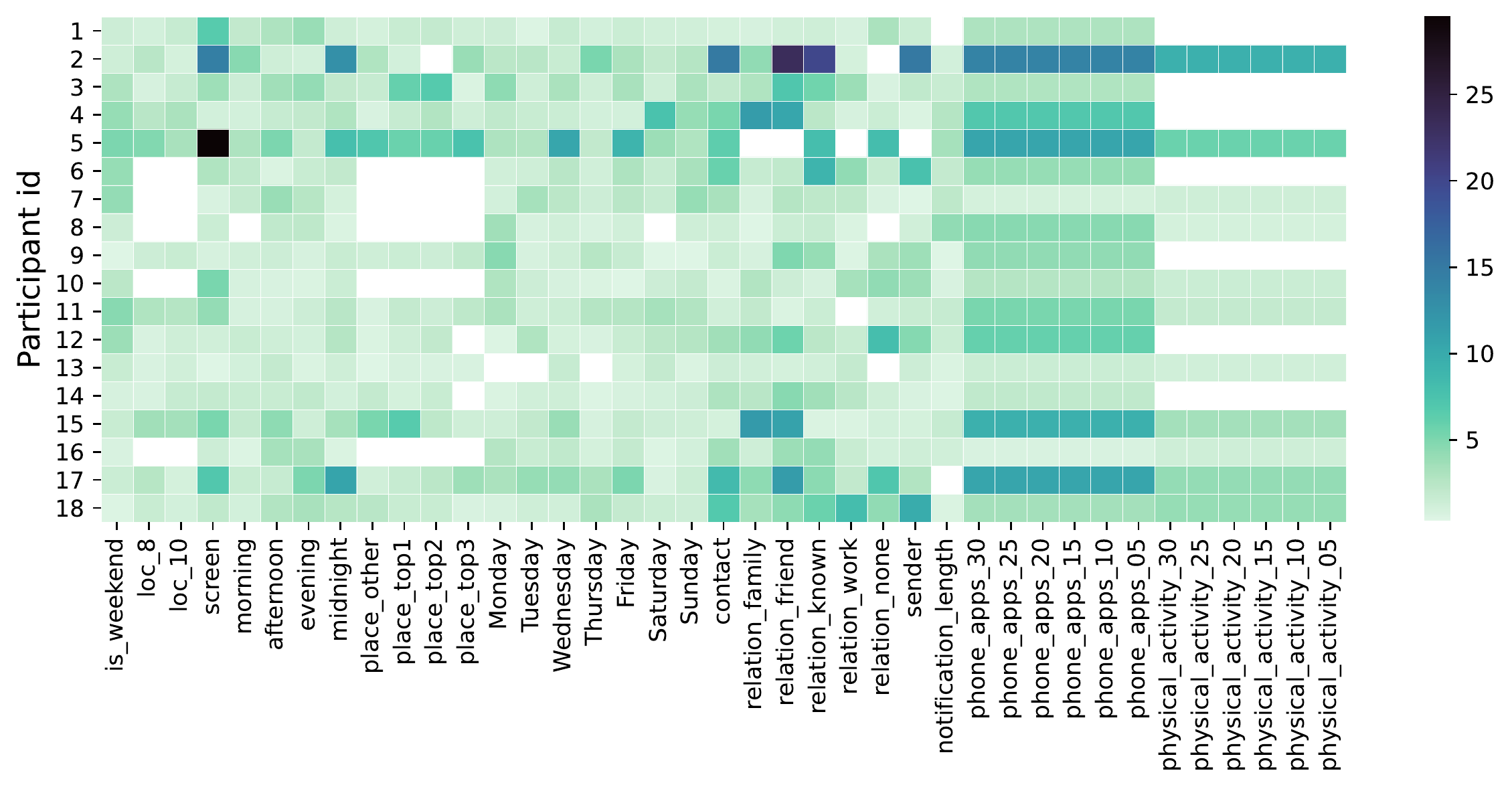}
    \caption{Feature importance for each participant in the prediction}
    \label{fig:im}
\end{figure}

Figure \ref{fig:im} shows the feature importance for each participant, which was calculated using f-regression score in the \textit{scikit-learn} python package. Higher values indicate more important features. Understanding the importance of a feature is significant in helping us better understand a problem and can lead to better prediction performance through feature selection. In Figure \ref{fig:im}, we can see obvious individual differences in feature importance for predicting notification response times. For example, the response time for some participants (e.g., P3, P5 and P15) was significantly affected by location, while some participants' (e.g., P12 and P13) were not affected by location. Many participants' response time  were influenced by the daytime, workday or not, screen status, relationship with senders or the number of apps used in the past 5, 10, 15, 20, 25, 30 minutes. The above phenomena are in line with our daily experience and may be due to the various personalities or usage habits of mobile users \cite{westermann2015assessing, gao2019predicting}.


\subsection{Impact of Mood-related Features}
We also explored the impact of mood-related features on predicting notification response times. The mood-related features were divided into two groups: ESM features and E4 features. For ESM features, we mainly focused on the perceived arousal and valence, based on ESM questionnaires. For E4 features, we mainly focused on the features extracted from physiological signals (i.e., EDA, HRV and ACC) from the E4 wristbands.

\textit{ESM features}. We built regression models with two different sets of features: (1) mobile features and ESM features; (2) mobile features only. Since we only had limited number of ESM responses, we removed the data instances without corresponding arousal and valence values. To achieve a fair comparison, we used the exact same rows of data (8408 data instances) in each of the above two different models. The results of the experiment showed that all the regression models using the second set of features had higher MAE and RMSE values than those using the first set of features, where the MAE/RMSE of baseline models are exactly the same. The findings indicated that the ESM features improved the prediction performance of the model for notification response times. 

\textit{E4 features}. To study the impact of the E4 features, we built regression models with two different sets of features: (1) mobile features and E4 features; (2) mobile features only. After removing the NaN values in the whole dataset, 1491 rows of data instances remained,  which were used to build the regression models using the two sets of features, as mentioned above. The results of the experiment showed that most of the regression models (except \textit{Bayesian} regressor) achieved better prediction performance with the first set of features, i.e. mobile features and E4 features. A possible reason may be the small number of E4 data instances, e.g. participant P11 only had 19 rows of data instances and P9 only had 27 rows of data instances, which makes it difficult to make meaningful predictions.

\section{Implications and Limitations}
\label{sec:implications}
\subsection{Implications}
This research addressed the relationship between mood and interruptiblity and investigated the possibility of automatically predicting notification response times and actions based on users' moods. Our research also provides opportunities for the future design of intelligent notification management systems for the mobile or desktop devices, which could benefit the wellbeing and productivity of users.
In our paper, we analysed the impact of mood, as measured by ESM questionnaires, and physiological data, as measured by E4 wristbands, on notification response times. We found that affective data can help to improve regression models to assist in the handling of smartphone notifications.

\subsection{Limitations}

\textbf{ESM data:}
One limitation of our study is that some data, such as mood, was gathered using an ESM questionnaire pushed either every 90 minutes or after the user had been using their smartphone usage for 10 min. This kind of questionnaire must be seen as an interruption itself. In addition, the questionnaire popped up on, the smartphone as a notification, which may have caused the participants to interact with their smartphones more often than they would normally have. However, this method of data collection is very common in the field of interruption management, and as the data were used to develop the  individual regression models, we believe that these initial results are valuable for further research. We are aware that a follow-up in-the-wild study is needed to validate the models developed.

\textbf{Mood:} Another limitation is the use of the ESM questionnaire to capture the participant's mood. It is important to note that many people struggle to identify or name their moods correctly \cite{kreibig2010}, and the reliability of self-report data can be influenced by various response biases \cite{gao2021investigating}. To compensate for this weakness, we added physiological signals to the \ac{sam} data, which also conveys information about human affective states. Even though these are not free of external influences (e.g. external temperature and physical movement), they form a basis for the research in combination with the ESM data.

\textbf{Data Distribution:}
There was minimal diversity in terms of age and gender, and there were only a small number of participants. In particular, the number of participants wearing the E4 wristband needs to be increased in future research to reduce the bias. \hl{Additionally, no application for iPhones or other Smartphone OS-Systems than Android was implemented. Likewise,} the data were very unbalanced because of the number of different apps used by each participant and the number of notifications (see Figure ). There was significant variation in how the subjects behaved and the apps that they used. Some users interacted frequently with many apps, while some users frequently interacted with a few apps and rarely with many other apps. These factors mainly influenced the results of the regression analysis, making it almost impossible to create a generalised model. After pre-processing, we also recognised that for some participants the quantity of data recorded was very low. \hl{This problem can be addressed in future work by measuring more participants and data. First, the users could be clustered according to their app usage behavior patterns. After that, response time prediction models created for the different behavior patterns could follow. This process would also enable a cold start for new users.}



\section{Conclusion}
\label{sec:conclusion}
Understanding the notification response behaviour of users is of vital importance for developing the next-generation mobile management system to improve users' productivity and well-being in daily lives. In this research, we predict notification response time by understanding people's mobile usage behaviours, mood, and physiological patterns. We have conducted an \textit{in-the-wild} study of more than $18$ participants with mobile devices and wearables in a five-week data collection. We develop multiple regression models to predict the notification response time for each participant. The experimental results show that the proposed model achieves higher prediction performance than all the baselines. We find that the use of mood data in the form of ESM and physiological signals (e.g., EDA and HRV) improves the prediction significantly. 
In addition, we identify the most significant features affecting the prediction of notification response time for each participant. Further, we discuss various factors affecting the prediction performance such as the individual differences and categories of applications.
Overall, our research showed that the notification response time can be predicted accurately using smartphone data (such as location, application usage,etc.), and the prediction performance can be significantly improved by utilizing mood-related features from ESM data or physiological signals. This result is a significant step toward achieving an attention management system that combines human well-being and behavior.

\begin{acks}
This research was partly conducted as part of RoboTrust, a project of the Centre Responsible Digitality.
\end{acks}

\bibliographystyle{ACM-Reference-Format}
\bibliography{main}

\appendix
\end{document}